# Empirically-based Multibody Dynamics for Modeling the Human Body Musculoskeletal System

**Hossein Ehsani, PhD**

*Neuroscience and Cognitive Science Program, University of Maryland*

*Kinesiology Department, University of Maryland*

## Prologue

This work serves as an English translation of a portion of my PhD dissertation, originally published in 2016 in Farsi. The translated document primarily encompasses the majority of Chapter 3 and sections of Chapter 4 from the original work. Chapter 3 of the original dissertation focused on the development of the methodology, while Chapter 4 presented the simulation results. It should be noted that the original document was published seven years ago, and although there may have been advancements in the field during this time, this translation remains faithful to the original text without incorporating any new materials or alterations.

While certain parts of this chapter were previously published as manuscripts in scientific journals, it is important to highlight that the main methodology for deriving the equations of motion, which was previously unpublished and not in English, is now being made publicly available through this translation. In this regard, appropriate citations have been included to reference the aforementioned papers.

If you have any questions or requests, such as code sharing, regarding this manuscript, please feel free to contact me at hehsani@umd.edu or [hossein.ehsani.bme@gmail.com](mailto:hossein.ehsani.bme@gmail.com).

Of note, the original dissertation was co-advised by Dr. Mostafa Rostami and Dr. Mohamad Parnianpour at the Amirkabir University of Technology. The citation for the original dissertation reads as follows:

Ehsani, H., "Developing an EMG-driven model for the human shoulder considering the shoulder rhythm", PhD Dissertation in Farsi, Amirkabir University of Technology, 2016.

If you use this translation in your research, please make sure that you cite the English translation available in arxiv. In addition, citing the original dissertation is encouraged but not required since you may not have access to it.

# Abstract[1]

This study introduces a novel approach for deriving the governing equations of the musculoskeletal system in the human body. The proposed formalism offers a framework to effectively incorporate the kinematic characteristics of biological joints and the complexities of kinematic chains into the differential equations of motion. This approach, known as "Empirically-based multibody dynamics," relies on experimental data pertaining to the skeletal system.

To establish the formulations, a novel calculus of matrix-valued functions is employed. In contrast to alternative methods that utilize Differential-Algebraic Equations (DAEs), the current approach provides the governing equations of the musculoskeletal system in the form of ordinary differential equations, simplifying the computational process. This formalism shall be employed to simulate a generic constrained multibody system with holonomic constraints.

The proposed formalism is applied to simulate various conventional multibody systems, including the four-bar linkage, five-bar linkage, and Andrew's squeezer mechanism. Additionally, the human shoulder mechanism is simulated, incorporating empirically-derived constraints to capture the shoulder rhythm. Furthermore, a novel method for deriving a closed-form formula for the moment arm matrix of a general musculotendon model is presented in this work.

---

[1] This abstract has not been translated from the original document and reflects a summary of the current document.

# 1- Introduction[2]

The musculoskeletal system of the human body can be represented as a series of interconnected rigid bodies. These bodies, which correspond to bones, are connected through biological joints, and they may form intricate kinematic chains. While it is possible to simulate these joints using ideal mechanical joints, such as a revolute joint for the knee, a universal joint for the elbow, and a ball-and-socket joint for the hip, and disregard the complexities of the kinematic chains, such as the shoulder rhythm, such paradigms raise doubts about the accuracy and realism of these models.

The kinematic characteristics of biological joints can be determined through empirical means. For instance, in the tibiofemoral joint, the translations between the proximal end of the tibia and the distal end of the femur have been mathematically described as functions of the knee flexion angle (Nisell, Nemeth et al. 1986, Yamaguchi and Zajac 1989, Delp, Loan et al. 1990). Additionally, the orientation and position of the patellofemoral joint have been expressed as smooth functions of the knee flexion angle (van Eijden, de Boer et al. 1985). Researchers have employed experimental methods to quantify the migration of the instant center of rotation in the elbow joint during flexion (Chao and Morrey 1978). Furthermore, it has been demonstrated that the ankle complex exhibits contingent movement. Within this complex, the talocrural joint possesses one degree of freedom, following a predetermined path (Dempster 1955, Wong, Kim et al. 2005).

---

[2] Portions of this introduction, with certain modifications, have been previously published in Ehsani, H., M. Poursina, M. Rostami, A. Mousavi, M. Parnianpour and K. Khalaf (2019). "Efficient embedding of empirically-derived constraints in the ODE formulation of multibody systems: Application to the human body musculoskeletal system." <u>Mechanism and Machine Theory</u> **133**: 673-690.

Empirical approaches have also played a crucial role in uncovering the kinematic relationships among the segments that constitute a complex kinematic chain. For instance, previous studies have delved into the shoulder rhythm, which involves the interdependence of clavicle and scapula orientations with respect to two generalized coordinates, namely the plane of elevation and elevation angle (Hogfors, Sigholm et al. 1987, Hogfors, Peterson et al. 1991, de Groot 1997, de Groot and Brand 2001). This interdependency arises from the glenohumeral rotations and the closed-chain mechanism involving the thorax, clavicle, and scapula (de Groot and Brand 2001). Alongside the shoulder complex, researchers have also experimentally investigated the coupled movement of the spine and pelvis in the sagittal plane, known as the lumbar-pelvic rhythm. In this context, Anderson et al. (Anderson, Chaffin et al. 1985) have examined the dependency of sacral rotation on torso rotation and knee flexion angle.

In order to develop realistic subject-specific skeletal models of the human body, it is crucial to incorporate the experimental information mentioned earlier into the governing equations of the system. Previous studies in this field can be classified into three groups.

The first group focuses on examining static or quasi-static situations. For instance, Yamaguch and Zajac utilized reported experimental data for the tibiofemoral and patellofemoral joints to estimate the moment arms of the quadriceps muscles in a mathematical model of the human knee joint (Yamaguchi and Zajac 1989). Similarly, Karlsson and Peterson (Karlsson and Peterson 1992) used experimental data reported by Hogfors et al. (Hogfors, Peterson et al. 1991) to present a model for force prediction in the human shoulder during static poses, considering the shoulder rhythm.

In the second group, differential equations of motion are derived while sacrificing some fidelity to the experimental observations. For example, Anderson and Pandy represented the human knee as a single degree of freedom revolute joint (Anderson and Pandy 1999), but incorporated measured values of the quadriceps moment arm from the study of Spoor and van Leeuwen (Spoor and Van Leeuwen 1992) to account for the role of the patella. Holzbaur et al. (Holzbaur, Murray et al. 2005) presented a 3-dimensional musculoskeletal model for the upper extremity, simplifying the experimental data on shoulder rhythm from de Groot and Brand (de Groot and Brand 2001). Similarly, Chadwick et al. (Chadwick, Blana et al. 2009) introduced the assumption of a fixed scapula into their 3-dimensional musculoskeletal model for real-time dynamic simulations of arm movements to avoid complexities associated with the shoulder rhythm.

In the third group, experimental observations are incorporated into the governing equations of the skeletal system through algebraic constraints, resulting in Differential-Algebraic Equations (DAEs). This approach can be seen in studies by Moissenet et al., Dumas et al., and Ribeiro for modeling the lower extremity (Dumas, Moissenet et al. 2012, Moissenet, Chèze et al. 2012, Ribeiro, Rasmussen et al. 2012), and by Garner and Pandy (Garner and Pandy 1999), Quental et al. (Quental, Folgado et al. 2012), and De Sapio (De Sapio, Holzbaur et al. 2006) for modeling the human shoulder. However, solving the derived DAEs numerically can pose challenges. Constraint violation becomes a significant issue over time due to error propagation and accumulation, requiring stabilization techniques (Baumgarte 1972, Koul, Shah et al. 2014). While these techniques are applicable, selecting appropriate parameters for the stabilizing equations is a critical challenge (Koul, Shah et al. 2014). Additionally, this method is computationally demanding, especially for tracking or dynamic optimization problems.

Given the unique structure of the human body's skeletal system, which includes biological joints and complex kinematic chains, and considering the current state of the field, it is evident that the biomechanical community would greatly benefit from novel approaches in deriving the governing equations for the skeletal system. In this study, I propose a computationally efficient method for deriving these equations specifically for the human body. With the primary objective of developing a subject-specific, EMG-driven model of the human shoulder, it is crucial to first determine the shoulder rhythm empirically for each participant and incorporate it into the governing equations. Accurately describing this rhythm is essential for determining the moment arm and length of the musculotendon units spanning the shoulder. However, representing this rhythm as a set of algebraic equations alongside the differential equations of motion would render the problem computationally infeasible, particularly in the context of parameter and structure identification within the modeling process.

In this chapter, I present a novel and computationally efficient method for deriving the governing equations of the skeletal system in the human body. This approach, referred to as "Empirically-based multibody dynamics," relies heavily on experimental information about the skeletal system. Unlike the methods described in the third group, this method expresses the governing equations using ordinary differential equations exclusively. To derive these equations, we employ the calculus of matrix-valued functions.

While previous studies by Vetter (Vetter 1970, Vetter 1973, Vetter 1975) and Brewer (Brewer 1978) have established theorems for obtaining derivatives of matrix-valued functions with respect to other matrices, I have developed a new definition of the partial derivative of a matrix-valued function with respect to a vector. In this dissertation, I refrain from using these theorems and instead introduce this new definition, presenting new theorems and properties in this chapter. With this new calculus for matrix-valued functions, I will proceed to develop closed-form formulations for both kinematics and dynamics of constrained multibody systems, integrating algebraic constraints into the equations.

The chapter begins by introducing the definition of the partial derivative of a matrix-valued function with respect to a vector. Subsequently, several theorems and properties resulting from this calculus are presented and proven. The next section delves into kinematic analysis of constrained multibody systems using the newly developed calculus. Specifically, the kinematic analysis of two different constrained multibody systems is examined, and the results are compared with traditional methods commonly employed for such analysis. Following the kinematic analysis, the chapter proceeds to discuss modeling the shoulder rhythm from a kinematic perspective. Spatial orientations of the clavicle, scapula, and humerus with respect to an inertial reference frame are represented, and the collection of kinematic data necessary for developing a subject-specific model of the shoulder rhythm is explained. In the subsequent section, a novel method for deriving the governing equation of a constrained multibody system is presented. This method utilizes the formulation derived for kinematic analysis and incorporates the virtual work principle. Additionally, the concept of the generalized musculotendon line of action is defined, leading to the development of a closed-form formula for the moment arm matrix of a musculoskeletal model. Finally, the proposed algorithms and methods are evaluated by considering three different problems. The validity and effectiveness of the proposed approaches are discussed in light of these evaluations.

## 2- Calculus of Matrix-valued Functions[3]

Before we delve into the definition of the partial derivative of a matrix-valued function with respect to a vector, I will review some fundamental concepts. In this manuscript, an elementary matrix is denoted as $\mathcal{E}_{ik}^{m \times n}$. This matrix belongs to $\mathbb{R}^{m \times n}$, and all its elements are zero, except for the element located at the $i^{th}$ row and $k^{th}$ column, which is equal to 1. By utilizing elementary matrices, an arbitrary matrix $\mathbf{A} \in \mathbb{R}^{m \times n}$ can be represented as follows:

$$\mathbf{A} = \sum_{i=1}^{m} \sum_{k=1}^{n} A_{ik} \mathcal{E}_{ik}^{m \times n} \qquad (1)$$

where $A_{ik}$ is an element of matrix $\mathbf{A}$ located at the $i^{th}$ row and $k^{th}$ column.

Identity matrices are shown as $\mathbf{I}_m$ where the subscript $m$ indicates that $\mathbf{I}_m \in \mathbb{R}^{m \times m}$. Furthermore, using elementary matrices, a permutation matrix $\mathbf{U}_{m,n} \in \mathbb{R}^{mn \times mn}$, is defined as:

$$\mathbf{U}_{m,n} = \sum_{i=1}^{m} \sum_{k=1}^{n} \mathcal{E}_{ik}^{m \times n} \otimes \mathcal{E}_{ki}^{n \times m} . \qquad (2)$$

---

[3] Definitions and theorems of this section, except for Theorem 3, were first published in Ehsani, H., M. Rostami and M. Parnianpour (2016). "A closed-form formula for the moment arm matrix of a general musculoskeletal model with considering joint constraint and motion rhythm." <u>Multibody System Dynamics</u> **36**: 377-403.

**Definition 1: Partial derivative of a matrix-valued function with respect to a vector**

Let $\mathbf{A} \in \mathbb{R}^{m \times n}$ and $\mathbf{q} \in \mathbb{R}^{r}$. Assuming that the partial derivative of all elements of this matrix-valued function respect to this vector exists, the partial derivative of $\mathbf{A}$ with respect to $\mathbf{q}$ is defined as follows:

$$\frac{\partial \mathbf{A}}{\partial \mathbf{q}} = \sum_{i=1}^{m} \sum_{k=1}^{n} \mathcal{E}_{ik}^{m \times n} \otimes \frac{\partial A_{ik}}{\partial \mathbf{q}} \qquad (3)$$

where $\otimes$ shows the Kronecker multiplication for matrices. According to this definition one should note that $\frac{\partial \mathbf{A}}{\partial \mathbf{q}} \in \mathbb{R}^{m\,r \times n}$.

In the subsequent discussion, I will present four theorems to elaborate on the properties of this definition. It is important to note that all matrices involved in these theorems adhere to the conditions stated in Definition 1. Throughout the proofs that follow, I will make use of the properties of Kronecker multiplication. A detailed description and proofs of these properties can be found in references on linear algebra, particularly in (Neudecker 1969, Magnus and Neudecker 1979, Henderson and Searle 1981, Loan 2000).

**Theorem 1: Partial derivative of matrix-valued function multiplication**

For two arbitrary matrix-valued functions $\mathbf{A} \in \mathbb{R}^{m \times n}$ and $\mathbf{B} \in \mathbb{R}^{n \times t}$, the derivative of their multiplication with respect to an arbitrary vector $\mathbf{q} \in \mathbb{R}^{r}$ is

$$\frac{\partial (\mathbf{AB})}{\partial \mathbf{q}} = \frac{\partial \mathbf{A}}{\partial \mathbf{q}} \mathbf{B} + \left( \mathbf{A} \otimes \mathbf{I}_r \right) \frac{\partial \mathbf{B}}{\partial \mathbf{q}}. \qquad (4)$$

**Proof**

Using the definition of matrix multiplication and Definition 1, one may have

$$\frac{\partial (\mathbf{AB})}{\partial \mathbf{q}} = \sum_{i=1}^{m} \sum_{k=1}^{t} \mathcal{E}_{ik}^{m \times t} \otimes \sum_{j=1}^{n} \left( \frac{\partial A_{ij}}{\partial \mathbf{q}} B_{jk} + A_{ij} \frac{\partial B_{jk}}{\partial \mathbf{q}} \right), \tag{5}$$

in other words,

$$\frac{\partial (\mathbf{AB})}{\partial \mathbf{q}} = \sum_{i=1}^{m} \sum_{k=1}^{t} \sum_{j=1}^{n} \left( B_{jk} \mathcal{E}_{ik}^{m \times t} \right) \otimes \frac{\partial A_{ij}}{\partial \mathbf{q}} + \sum_{i=1}^{m} \sum_{k=1}^{t} \sum_{j=1}^{n} \left( A_{ij} \mathcal{E}_{ik}^{m \times t} \right) \otimes \frac{\partial B_{jk}}{\partial \mathbf{q}}. \tag{6}$$

Using the definition of an elementary matrix yields

$$B_{jk} \mathcal{E}_{ik}^{m \times t} = \mathcal{E}_{ij}^{m \times n} \mathbf{B} \mathcal{E}_{kk}^{t \times t} \tag{7}$$

and,

$$A_{ij} \mathcal{E}_{ik}^{m \times t} = \mathcal{E}_{ii}^{m \times m} \mathbf{A} \mathcal{E}_{jk}^{n \times t}. \tag{8}$$

Using Eq. 7 and Eq. 8, one may rewrite Eq. 6 as follows:

$$\frac{\partial (\mathbf{AB})}{\partial \mathbf{q}} = \sum_{i=1}^{m} \sum_{k=1}^{t} \sum_{j=1}^{n} \left( \mathcal{E}_{ij}^{m \times n} \mathbf{B} \mathcal{E}_{kk}^{t \times t} \right) \otimes \frac{\partial A_{ij}}{\partial \mathbf{q}} + \sum_{i=1}^{m} \sum_{k=1}^{t} \sum_{j=1}^{n} \left( \mathcal{E}_{ii}^{m \times m} \mathbf{A} \mathcal{E}_{jk}^{n \times t} \right) \otimes \frac{\partial B_{jk}}{\partial \mathbf{q}}. \tag{9}$$

Employing the mixed-product property of the Kronecker multiplication, this equation is simplified

$$\begin{aligned}\frac{\partial (\mathbf{AB})}{\partial \mathbf{q}} &= \sum_{i=1}^{m} \sum_{j=1}^{n} \left( \mathcal{E}_{ij}^{m \times n} \otimes \frac{\partial A_{ij}}{\partial \mathbf{q}} \right) \left( \mathbf{B} \sum_{k=1}^{t} \mathcal{E}_{kk}^{t \times t} \otimes 1 \right) \\ &+ \left( \left( \sum_{i=1}^{m} \mathcal{E}_{ii}^{m \times m} \mathbf{A} \right) \otimes \mathbf{I}_r \right) \sum_{k=1}^{t} \sum_{j=1}^{n} \left( \mathcal{E}_{jk}^{n \times t} \otimes \frac{\partial B_{jk}}{\partial \mathbf{q}} \right) \end{aligned}. \tag{10}$$

Since $\mathbf{I}_m = \sum_{i=1}^{m} \mathcal{E}_{ii}^{m \times m}$ and $\mathbf{I}_t = \sum_{k=1}^{t} \mathcal{E}_{kk}^{t \times t}$, the theorem is proven.

**Theorem 2: Partial derivative of the multiplication of a scalar function and a matrix-valued function**

Let the scalar function $a$ be smooth over an arbitrary vector $\mathbf{q} \in \mathbb{R}^r$. Thus, the partial derivative of the multiplication of an arbitrary matrix-valued function $\mathbf{A} \in \mathbb{R}^{m \times n}$ and this scalar function with respect to $\mathbf{q}$ is given by

$$\frac{\partial (a\mathbf{A})}{\partial \mathbf{q}} = \mathbf{A} \otimes \frac{\partial a}{\partial \mathbf{q}} + a \frac{\partial \mathbf{A}}{\partial \mathbf{q}}. \tag{11}$$

**Proof**

Using Definition 1 yields

$$\begin{aligned} \frac{\partial (a\mathbf{A})}{\partial \mathbf{q}} &= \sum_{i=1}^{m} \sum_{k=1}^{n} \mathcal{E}_{ik}^{m \times n} \otimes \frac{\partial (a A_{ik})}{\partial \mathbf{q}} \\ &= \sum_{i=1}^{m} \sum_{k=1}^{n} \mathcal{E}_{ik}^{m \times n} \otimes \left( \frac{\partial a}{\partial \mathbf{q}} A_{ik} + a \frac{\partial A_{ik}}{\partial \mathbf{q}} \right) \end{aligned}. \tag{12}$$

Employing the mixed-product property of the Kronecker multiplication, one may infer

$$\frac{\partial (a\mathbf{A})}{\partial \mathbf{q}} = \sum_{i=1}^{m} \sum_{k=1}^{n} \left( \mathcal{E}_{ik}^{m \times n} A_{ik} \right) \otimes \frac{\partial a}{\partial \mathbf{q}} + a \sum_{i=1}^{m} \mathcal{E}_{ik}^{m \times n} \otimes \frac{\partial A_{ik}}{\partial \mathbf{q}}, \tag{13}$$

which proves the theorem. Of note, this theorem could be considered as a special case of Theorem 1.

**Theorem 3: Partial derivative of the Kronecker multiplication of two matrix-valued functions**

For two arbitrary matrix-valued functions $\mathbf{A} \in \mathbb{R}^{m \times n}$ and $\mathbf{C} \in \mathbb{R}^{s \times p}$, the derivative of their Kronecker multiplication with respect to an arbitrary vector $\mathbf{q} \in \mathbb{R}^r$ is

$$\frac{\partial(\mathbf{A}\otimes\mathbf{C})}{\partial\mathbf{q}}=\left(\mathbf{I}_m\otimes\mathbf{U}_{s,r}\right)\left(\frac{\partial\mathbf{A}}{\partial\mathbf{q}}\otimes\mathbf{C}\right)+\mathbf{A}\otimes\frac{\partial\mathbf{C}}{\partial\mathbf{q}}. \tag{14}$$

**Proof**

Since $\mathbf{A}=\sum_{i=1}^{m}\sum_{k=1}^{n}A_{ik}\mathcal{E}_{ik}^{m\times n}$ and $\mathbf{C}=\sum_{j=1}^{s}\sum_{l=1}^{p}C_{jl}\mathcal{E}_{jl}^{s\times p}$, $\mathbf{A}\otimes\mathbf{C}$ is

$$\mathbf{A}\otimes\mathbf{C}=\sum_{i=1}^{m}\sum_{k=1}^{n}\sum_{j=1}^{s}\sum_{l=1}^{p}\left(\mathcal{E}_{ik}^{m\times n}\otimes\mathcal{E}_{jl}^{s\times p}\right)A_{ik}C_{jl}. \tag{15}$$

Taking the partial derivative with respect to $\mathbf{q}$ on both sides of this equation results in

$$\frac{\partial(\mathbf{A}\otimes\mathbf{C})}{\partial\mathbf{q}}=\sum_{i=1}^{m}\sum_{k=1}^{n}\sum_{j=1}^{s}\sum_{l=1}^{p}\left(\mathcal{E}_{ik}^{m\times n}\otimes\mathcal{E}_{jl}^{s\times p}\right)\otimes\frac{\partial\left(A_{ik}C_{jl}\right)}{\partial\mathbf{q}}. \tag{16}$$

Of note, to find this equation, I have used Theorem 1 and the fact that $\frac{\partial}{\partial\mathbf{q}}\left(\mathcal{E}_{ik}^{m\times n}\otimes\mathcal{E}_{jl}^{s\times p}\right)$ vanishes. The right-hand side of Eq. 16 can be expanded as

$$\frac{\partial(\mathbf{A}\otimes\mathbf{C})}{\partial\mathbf{q}}=\sum_{i=1}^{m}\sum_{k=1}^{n}\sum_{j=1}^{s}\sum_{l=1}^{p}\left(\mathcal{E}_{ik}^{m\times n}\otimes\mathcal{E}_{jl}^{s\times p}\right)\otimes\left(\frac{\partial A_{ik}}{\partial\mathbf{q}}C_{jl}+A_{ik}\frac{\partial C_{jl}}{\partial\mathbf{q}}\right); \tag{17}$$

in other words,

$$\frac{\partial(\mathbf{A}\otimes\mathbf{C})}{\partial\mathbf{q}}=\underbrace{\sum_{i=1}^{m}\sum_{k=1}^{n}\sum_{j=1}^{s}\sum_{l=1}^{p}\left(\mathcal{E}_{ik}^{m\times n}\otimes C_{jl}\mathcal{E}_{jl}^{s\times p}\right)\otimes\frac{\partial A_{ik}}{\partial\mathbf{q}}}_{1^{\text{st}}\text{ part}}+\underbrace{\sum_{i=1}^{m}\sum_{k=1}^{n}\sum_{j=1}^{s}\sum_{l=1}^{p}\left(A_{ik}\mathcal{E}_{ik}^{m\times n}\otimes\mathcal{E}_{jl}^{s\times p}\right)\otimes\frac{\partial C_{jl}}{\partial\mathbf{q}}}_{2^{\text{nd}}\text{ part}}. \tag{18}$$

To better simplify Eq. 18, it has been divided into two parts. Using the associativity property of the Kronecker multiplication on the 1st part results in

$$1^{st} \text{ part} = \sum_{i=1}^{m}\sum_{k=1}^{n}\sum_{j=1}^{s}\sum_{l=1}^{p} \mathcal{E}_{ik}^{m\times n} \otimes \left( C_{jl}\, \mathcal{E}_{jl}^{s\times p} \otimes \frac{\partial A_{ik}}{\partial \mathbf{q}} \right) \qquad (19)$$

Next, employing the commutative rule of the Kronecker multiplication (Loan 2000) yields

$$1^{st} \text{ part} = \sum_{i=1}^{m}\sum_{k=1}^{n}\sum_{j=1}^{s}\sum_{l=1}^{p} \mathcal{E}_{ik}^{m\times n} \otimes \left[ \mathbf{U}_{s,r} \left( \frac{\partial A_{ik}}{\partial \mathbf{q}} \otimes C_{jl}\, \mathcal{E}_{jl}^{s\times p} \right) \right]. \qquad (20)$$

Applying the mixed-product property of the Kronecker multiplication on Eq. 20 would lead to the following:

$$1^{st} \text{ part} = \sum_{i=1}^{m}\sum_{k=1}^{n}\sum_{j=1}^{s}\sum_{l=1}^{p} \left( \mathbf{I}_m \otimes \mathbf{U}_{s,r} \right) \left[ \mathcal{E}_{ik}^{m\times n} \otimes \left( \frac{\partial A_{ik}}{\partial \mathbf{q}} \otimes C_{jl}\, \mathcal{E}_{jl}^{s\times p} \right) \right]. \qquad (21)$$

Therefore,

$$1^{st} \text{ part} = \left( \mathbf{I}_m \otimes \mathbf{U}_{s,r} \right) \sum_{i=1}^{m}\sum_{k=1}^{n}\sum_{j=1}^{s}\sum_{l=1}^{p} \mathcal{E}_{ik}^{m\times n} \otimes \left( \frac{\partial A_{ik}}{\partial \mathbf{q}} \otimes C_{jl}\, \mathcal{E}_{jl}^{s\times p} \right). \qquad (22)$$

Since associativity holds in Kronecker multiplications, Eq. 22 would be simplified as

$$1^{st} \text{ part} = \left( \mathbf{I}_m \otimes \mathbf{U}_{s,r} \right) \sum_{i=1}^{m}\sum_{k=1}^{n}\sum_{j=1}^{s}\sum_{l=1}^{p} \left( \mathcal{E}_{ik}^{m\times n} \otimes \frac{\partial A_{ik}}{\partial \mathbf{q}} \right) \otimes C_{jl}\, \mathcal{E}_{jl}^{s\times p} ; \qquad (23)$$

consequently,

$$1^{st} \text{ part} = \left( \mathbf{I}_m \otimes \mathbf{U}_{s,r} \right) \left( \frac{\partial \mathbf{A}}{\partial \mathbf{q}} \otimes \mathbf{C} \right). \qquad (24)$$

To simplify the 2nd part, again the associativity of Kronecker multiplications is used

$$2^{nd} \text{ part} = \sum_{i=1}^{m}\sum_{k=1}^{n}\sum_{j=1}^{s}\sum_{l=1}^{p} A_{ik}\, \mathcal{E}_{ik}^{m\times n} \otimes \left( \mathcal{E}_{jl}^{s\times p} \otimes \frac{\partial C_{jl}}{\partial \mathbf{q}} \right); \qquad (25)$$

therefore,

$$2^{nd} \text{ part} = \mathbf{A} \otimes \frac{\partial \mathbf{C}}{\partial \mathbf{q}}. \tag{26}$$

Finally, by adding Eq. 24 and Eq. 26 the proof is complete.

**Theorem 4: The differential of a matrix-valued function**

The differential of an arbitrary matrix-valued function such as $\mathbf{A} \in \mathbb{R}^{m \times n}$ that is an explicit function of $\mathbf{q} \in \mathbb{R}^r$ is determined as

$$\text{a) } d\mathbf{A} = \left(\mathbf{I}_m \otimes d\mathbf{q}^T\right) \frac{\partial \mathbf{A}}{\partial \mathbf{q}}, \tag{27}$$

$$\text{b) } d\mathbf{A} = \left(\frac{\partial \mathbf{A}^T}{\partial \mathbf{q}}\right)^T \left(\mathbf{I}_n \otimes d\mathbf{q}\right). \tag{28}$$

**Proof**

a) Taking the differential of both sides of Eq. 1 yields

$$d\mathbf{A} = \sum_{i=1}^{m} \sum_{k=1}^{n} \mathcal{E}_{ik}^{m \times n} \, dA_{ik}. \tag{29}$$

Furthermore,

$$dA_{ik} = \sum_{j=1}^{r} \frac{\partial A_{ik}}{\partial q_j} dq_j = d\mathbf{q}^T \frac{\partial A_{ik}}{\partial \mathbf{q}}. \tag{30}$$

Substituting Eq. 30 into Eq. 29 results in

$$d\mathbf{A} = \sum_{i=1}^{m} \sum_{k=1}^{n} \mathcal{E}_{ik}^{m \times n} \left(d\mathbf{q}^T \frac{\partial A_{ik}}{\partial \mathbf{q}}\right). \tag{31}$$

Since $d\mathbf{q}^T \frac{\partial A_{ik}}{\partial \mathbf{q}}$ is a scalar, Eq. 31 is rewritten as follows:

$$d\mathbf{A} = \sum_{i=1}^{m}\sum_{k=1}^{n} \mathcal{E}_{ik}^{m\times n} \otimes \left( d\mathbf{q}^T \frac{\partial A_{ik}}{\partial \mathbf{q}} \right). \tag{32}$$

Using the mixed-product property of the Kronecker multiplication, one may draw the following conclusion:

$$d\mathbf{A} = \sum_{i=1}^{m}\sum_{k=1}^{n} \mathcal{E}_{ik}^{m\times n} \otimes \left( d\mathbf{q}^T \frac{\partial A_{ik}}{\partial \mathbf{q}} \right) \tag{33}$$

which proves the validity of part a.

b) This proof is essentially the same as part a. Rearranging the terms in Eq. 30 results in

$$dA_{ik} = \left( \frac{\partial A_{ik}}{\partial \mathbf{q}} \right)^T d\mathbf{q} ; \tag{34}$$

consequently,

$$d\mathbf{A} = \sum_{i=1}^{m}\sum_{k=1}^{n} \mathcal{E}_{ik}^{m\times n} \otimes \left( \left( \frac{\partial A_{ik}}{\partial \mathbf{q}} \right)^T d\mathbf{q} \right). \tag{35}$$

Similar to part a, this equation is factored as

$$d\mathbf{A} = \sum_{i=1}^{m}\sum_{k=1}^{n} \left( \mathcal{E}_{ik}^{m\times n} \otimes \left( \frac{\partial A_{ik}}{\partial \mathbf{q}} \right)^T \right) (\mathbf{I}_n \otimes d\mathbf{q}). \tag{36}$$

Using the fact that $\left( \mathcal{E}_{ik}^{m\times n} \right)^T = \mathcal{E}_{ki}^{n\times m}$, Eq. 36 is reshaped as

$$d\mathbf{A} = \sum_{i=1}^{m}\sum_{k=1}^{n} \left( \mathcal{E}_{ki}^{n\times m} \otimes \frac{\partial A_{ik}}{\partial \mathbf{q}} \right)^T (\mathbf{I}_n \otimes d\mathbf{q}). \tag{37}$$

This proves part b of this theorem.

**Remark**: Using Theorem 4, one can determine formulations to calculate $\frac{d\mathbf{A}}{dt}$ and $\delta \mathbf{A}$ in a similar fashion.

## 3- Kinematics

This section begins with a comprehensive review of the notations and nomenclature employed in the subsequent analysis. Subsequently, leveraging the established matrix calculus from the previous section, fundamental kinematic equations are derived. These formulations heavily rely on the gradient vector and Hessian matrix of the joint variables with respect to the generalized coordinate vector. Hence, in the final subsection, I will provide a detailed explanation of the methodology to derive these crucial quantities.

### 1-3- Assumptions and Notations

In this study, the multibody system consists of $n_b$ moving rigid bodies. These bodies are interconnected with each other or the ground using one degree of freedom joints, namely revolute and prismatic joints. Each articulation is associated with a single joint variable denoted as $f_i$ for $i = 1, 2, \ldots, n_j$, where $n_j$ represents the total number of joints present in the multibody. It is important to note that this assumption does not impose any limitations on the applicability of the proposed algorithm. Consequently, this methodology can be employed to simulate any constrained multibody system. In instances where a joint possesses more than one degree of freedom, such as a cylindrical joint, it can be transformed into a series of one degree of freedom joints interconnected by rigid bodies with zero inertia and length.

Due to kinematic constraints, the minimum number of generalized coordinates required to fully determine the kinematics of a constrained multibody system, denoted as $n_q$, is typically less than the total number of joint variables. Let $\mathbf{q} = \{q_1 \quad q_2 \quad \cdots \quad q_{n_q}\}^T$ represent the set of all necessary generalized coordinates for describing the constrained multibody system. Assuming that all constraints imposed on the system are holonomic, the joint variables can be expressed as smooth functions of the generalized coordinate vector, i.e., $f_i = f_i(\mathbf{q})$ for $i = 1, 2, \ldots, n_j$. For an arbitrary joint $i$, the gradient vector and Hessian matrix of the joint variable with respect to the generalized coordinate vector are represented by $\nabla f_i$ and $\mathbf{H}_i$, respectively. If the joint variables are determined empirically, the associated gradient vector and Hessian matrix can be easily obtained. However, we will address the problem of finding these quantities in a more general case, where a system of algebraic equations is imposed on the multibody system, in the subsequent subsections.

The rigid bodies within the system are numbered following the guidelines outlined in (Featherstone 2008). Consequently, the ground is designated as "0," while each moving body is assigned a unique identifier from the set $\{1, 2, \ldots, n_b\}$. The parent body's identifier can be determined using the $pr(.)$ operator, and the $Anc(.)$ operator is employed to obtain the identifiers of all the moving ancestors of a particular body.

Vector quantities, such as the position vector, will be displayed in boldface. Typically, vector quantities are denoted with four subscripts and superscripts. For example, the position vector of an arbitrary point $p$ located on body $i$ with respect to the $j^{th}$ coordinate frame is represented as ${}_{(k)}^{j}\mathbf{r}_{p}^{i}$. The left subscript indicates that this vector is expressed in the basis of the *k*th coordinate frame. To maintain brevity, if $i$ and $k$ are omitted, their values are assumed to be equal to $j$. If the point of interest is the body's center of mass, it will be denoted as $cm$ instead of $p$. Similarly, if the point is the origin of the coordinate frame associated with the body, it will be denoted as $o$. The same rules apply for velocity and acceleration vectors of an arbitrary point. For example, the velocity and acceleration vectors of the aforementioned point $p$ will be denoted as ${}_{(k)}^{j}\mathbf{v}_{p}^{i}$ and ${}_{(k)}^{j}\mathbf{a}_{p}^{i}$, respectively.

To represent the skew matrix corresponding to a vector, an overhead tilde is used. For instance, $\tilde{\mathbf{v}}$ represents the skew matrix associated with vector $\mathbf{V}$.

## 2-3- Fundamental Kinematic Equations

Each rigid body is equipped with a right-handed orthonormal local coordinate frame. As shown in Fig. 1, the local coordinate frame of an arbitrary body $i$ is determined in relation to its parent coordinate frame, i.e. $pr(i)$, through two unit vectors, i.e. $\mathbf{V}_{d_i}$ and $\mathbf{V}_{\theta_i}$, and two corresponding parameters, i.e. $d_i$ and $\theta_i$. To establish the local coordinate frame $i$, the parent coordinate frame is first translated by a distance $d_i$ along the unit vector $\mathbf{V}_{d_i}$. Subsequently, a rotation is performed around the unit vector $\mathbf{V}_{\theta_i}$ by an angle $\theta_i$. It is important to note that both unit vectors are defined within the local coordinate frame of $pr(i)$. In the case where the joint connecting body $i$ and its parent is revolute, $\theta_i$ represents the joint variable while $d_i$ remains constant. Conversely, if the joint is prismatic, the situation is reversed, with $\theta_i$ being constant and $d_i$ serving as the joint variable.

The position and orientation of the $i^{th}$ coordinate frame with respect to its parent are defined using a 4x4 homogeneous transformation matrix, i.e., $^{Pr(i)}_i\mathbf{A}_4$, which can be expressed as follows:

$$^{pr(i)}_i\mathbf{A}_4 = \left[\begin{array}{c|c} ^{pr(i)}_i\mathbf{A} & d_i\mathbf{V}_{d_i} \\ \hline \mathbf{0}^T_{3\times 1} & 1 \end{array}\right] \tag{38}$$

where $^{pr(i)}_i\mathbf{A}$ is the rotation matrix between these coordinate frames and is defined as

$$^{pr(i)}_i\mathbf{A} = \mathbf{I}_3 + \tilde{\mathbf{V}}_{\theta_i}\sin\theta_i + 2\left(\tilde{\mathbf{V}}_{\theta_i}\right)^2\sin^2\frac{\theta_i}{2}. \tag{39}$$

The position and orientation of the $i^{th}$ coordinate frame with respect to the inertial reference frame are also determined by the following formulation:

$$^0_i\mathbf{A}_4 = \prod_{j\in Anc(i)\cup\{i\}} {}^{pr(j)}_j\mathbf{A}_4. \tag{40}$$

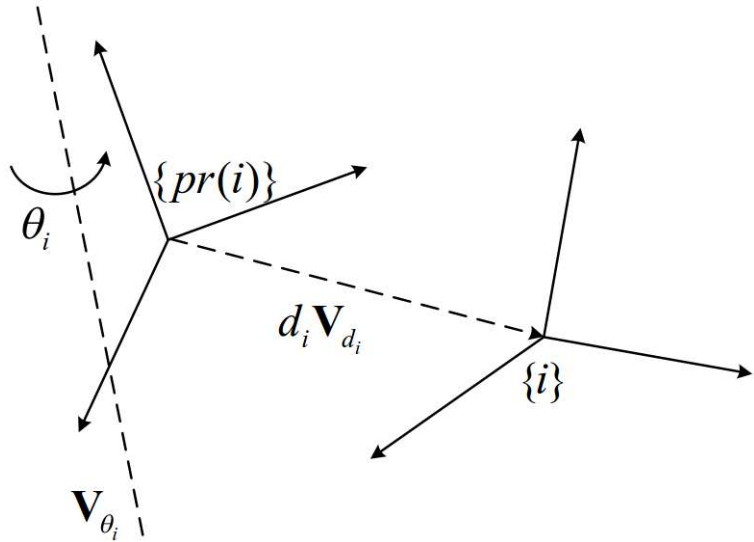

Figure 1 The definition of the $i^{th}$ local coordinate frame based on its parent. Refer to the text for more elaboration on the definition of the variables.

Accordingly, the position vector of an arbitrary point $p$ located in the $i^{th}$ body with respect to the inertial reference frame is obtained as

$$\left\{\begin{array}{c} ^{0}\mathbf{r}_{p}^{i} \\ \hline 1 \end{array}\right\} = {}^{0}_{i}\mathbf{A}_4 \left\{\begin{array}{c} ^{i}\mathbf{r}_{p} \\ \hline 1 \end{array}\right\}. \tag{41}$$

Taking the time derivative from both sides of this equation results in

$$\left\{\begin{array}{c} ^{0}\mathbf{v}_{p}^{i} \\ \hline 0 \end{array}\right\} = \frac{d\,{}^{0}_{i}\mathbf{A}_4}{dt} \left\{\begin{array}{c} ^{i}\mathbf{r}_{p} \\ \hline 1 \end{array}\right\}. \tag{42}$$

As ${}^{0}_{i}\mathbf{A}_4$ is a matrix-valued function of the generalized coordinate vector, to evaluate its time derivative, I utilize part b of Theorem 4 as follows:

$$\frac{d\,{}^{0}_{i}\mathbf{A}_4}{dt} = \left(\frac{\partial \mathbf{B}_i}{\partial \mathbf{q}}\right)^T (\mathbf{I}_4 \otimes \dot{\mathbf{q}}) \tag{43}$$

where for the sake of brevity in notations, I have used $\mathbf{B}_i = {}^{0}_{i}\mathbf{A}_4^T$. Substituting Eq. 43 into Eq. 42 and using the mixed-product property of the Kronecker multiplication yields

$$\left\{\begin{array}{c} ^{0}\mathbf{v}_{p}^{i} \\ \hline 0 \end{array}\right\} = \mathbf{D}_i^T \left(\left\{\begin{array}{c} ^{i}\mathbf{r}_{p} \\ \hline 1 \end{array}\right\} \otimes \dot{\mathbf{q}}\right) \tag{44}$$

where,

$$\mathbf{D}_i = \frac{\partial \mathbf{B}_i}{\partial \mathbf{q}}. \tag{45}$$

Of note, by utilizing Theorem 4 and its accompanying remark, one can derive an analogous formulation for the virtual position of point $p$, which is

$$\left\{ \frac{\delta^0 \mathbf{r}_p^i}{0} \right\} = \mathbf{D}_i^T \left( \left\{ \frac{^i \mathbf{r}_p}{1} \right\} \otimes \delta \mathbf{q} \right). \tag{46}$$

To derive a formula for the acceleration of point $p$, I shall take the time derivative of both sides of Eq. 44. Hence,

$$\left\{ \frac{^0 \mathbf{a}_p^i}{0} \right\} = \frac{d}{dt} \left( \left( \frac{\partial \mathbf{B}_i}{\partial \mathbf{q}} \right)^T \left( \left\{ \frac{^i \mathbf{r}_p}{1} \right\} \otimes \dot{\mathbf{q}} \right) \right). \tag{47}$$

To simplify the right-hand side of this equation, two separate parts are considered

$$\left\{ \frac{^0 \mathbf{a}_p^i}{0} \right\} = \underbrace{\frac{d}{dt} \left( \frac{\partial \mathbf{B}_i}{\partial \mathbf{q}} \right)^T \left( \left\{ \frac{^i \mathbf{r}_p}{1} \right\} \otimes \dot{\mathbf{q}} \right)}_{1^{st}\ part} + \underbrace{\left( \frac{\partial \mathbf{B}_i}{\partial \mathbf{q}} \right)^T \frac{d}{dt} \left( \left\{ \frac{^i \mathbf{r}_p}{1} \right\} \otimes \dot{\mathbf{q}} \right)}_{2^{nd}\ part}. \tag{48}$$

I begin the simplification of the 1$^{st}$ part by deriving $\frac{d}{dt} \left( \frac{\partial \mathbf{B}_i}{\partial \mathbf{q}} \right)^T$. As $\frac{\partial \mathbf{B}_i}{\partial \mathbf{q}} \in \mathbb{R}^{4n_q \times 4}$, employing Theorem 4 to calculate this derivative reads

$$\frac{d}{dt} \left( \frac{\partial \mathbf{B}_i}{\partial \mathbf{q}} \right)^T = \left( \frac{\partial^2 \mathbf{B}_i}{\partial \mathbf{q}^2} \right)^T \left( \mathbf{I}_{4n_q} \otimes \dot{\mathbf{q}} \right); \tag{49}$$

therefore,

$$1^{st}\ part = \left( \frac{\partial^2 \mathbf{B}_i}{\partial \mathbf{q}^2} \right)^T \left( \mathbf{I}_{4n_q} \otimes \dot{\mathbf{q}} \right) \left( ^i \mathbf{r}_p \otimes \dot{\mathbf{q}} \right). \tag{50}$$

As $1^{st}\ part = \left( \frac{\partial^2 \mathbf{B}_i}{\partial \mathbf{q}^2} \right)^T \left( \mathbf{I}_{4n_q} \otimes \dot{\mathbf{q}} \right) \left( ^i \mathbf{r}_p \otimes \dot{\mathbf{q}} \right)$, Eq. 40 is reshaped as follows:

$$1^{st}\ part = \left( \frac{\partial^2 \mathbf{B}_i}{\partial \mathbf{q}^2} \right)^T \left[ \left( \mathbf{I}_4 \otimes \mathbf{Z} \right) \left( \left\{ \frac{^i \mathbf{r}_p}{1} \right\} \otimes \dot{\mathbf{q}} \right) \right] \tag{51}$$

where,

$$\mathbf{Z} = \mathbf{I}_{n_q} \otimes \dot{\mathbf{q}}.$$

Using the mixed-product property of the Kronecker multiplication, one can derive the following:

$$(\mathbf{I}_4 \otimes \mathbf{Z})\left(\left\{\begin{matrix}{}^i\mathbf{r}_p \\ 1\end{matrix}\right\} \otimes \dot{\mathbf{q}}\right) = \left\{\begin{matrix}{}^i\mathbf{r}_p \\ 1\end{matrix}\right\} \otimes (\mathbf{Z}\dot{\mathbf{q}}). \tag{52}$$

Furthermore,

$$\mathbf{Z}\dot{\mathbf{q}} = \left(\mathbf{I}_{n_q} \otimes \dot{\mathbf{q}}\right)(\dot{\mathbf{q}} \otimes 1) = \dot{\mathbf{q}} \otimes \dot{\mathbf{q}}. \tag{53}$$

Consequently,

$$(\mathbf{I}_4 \otimes \mathbf{Z})\left(\left\{\begin{matrix}{}^i\mathbf{r}_p \\ 1\end{matrix}\right\} \otimes \dot{\mathbf{q}}\right) = \left\{\begin{matrix}{}^i\mathbf{r}_p \\ 1\end{matrix}\right\} \otimes \dot{\mathbf{q}} \otimes \dot{\mathbf{q}}. \tag{54}$$

By substituting results of Eq. 54 into Eq. 51, the 1st part is simplified as follows:

$$1^{st}\ \text{part} = \left(\frac{\partial^2 \mathbf{B}_i}{\partial \mathbf{q}^2}\right)^T \left(\left\{\begin{matrix}{}^i\mathbf{r}_p \\ 1\end{matrix}\right\} \otimes \dot{\mathbf{q}}^{[2]}\right) \tag{55}$$

where $\dot{\mathbf{q}}^{[2]} = \dot{\mathbf{q}} \otimes \dot{\mathbf{q}}$.

The 2nd part of Eq. 48 is readily simplified as follows:

$$2^{nd}\ \text{part} = \left(\frac{\partial \mathbf{B}_i}{\partial \mathbf{q}}\right)^T \left(\left\{\begin{matrix}{}^i\mathbf{r}_p \\ 1\end{matrix}\right\} \otimes \ddot{\mathbf{q}}\right). \tag{56}$$

As a result, the following formulation is derived to determine the acceleration vector of an arbitrary point $p$ of the $i^{th}$ body:

$$\left\{\dfrac{^{0}\mathbf{a}_{p}^{i}}{0}\right\} = \left(\mathbf{D}'_{i}\right)^{T}\left(\left\{\dfrac{^{i}\mathbf{r}_{p}}{1}\right\}\otimes \dot{\mathbf{q}}^{[2]}\right) + \mathbf{D}_{i}^{T}\left(\left\{\dfrac{^{i}\mathbf{r}_{p}}{1}\right\}\otimes \ddot{\mathbf{q}}\right) \qquad (57)$$

where,

$$\mathbf{D}'_{i} = \dfrac{\partial^{2}\mathbf{B}_{i}}{\partial \mathbf{q}^{2}}. \qquad (58)$$

As indicated in Eq. 44 and Eq. 57, in order to calculate the velocity and acceleration vectors of an arbitrary point $p$ situated on body $i$, it is essential to determine the first and second partial derivatives of $\mathbf{B}_{i}$ with respect to the generalized coordinate vector. In the subsequent part of this section, I will establish the required formulations to derive these quantities[4].

Firstly, I will derive a closed-form formula for $\dfrac{\partial}{\partial \mathbf{q}}{}^{pr(i)}_{i}\mathbf{A}_{4}^{T}$. Of note,

$$^{pr(i)}_{i}\mathbf{A}_{4}^{T} = \left[\begin{array}{c|c}{}^{pr(i)}_{i}\mathbf{A}^{T} & \mathbf{0}_{3\times 1} \\ \hline d_{i}\,\mathbf{V}_{d_{i}}^{T} & 1\end{array}\right] \qquad (59)$$

where,

$$^{pr(i)}_{i}\mathbf{A}^{T} = \mathbf{I}_{3} + \tilde{\mathbf{V}}_{\theta_{i}}^{T}\sin\theta_{i} + 2\left(\tilde{\mathbf{V}}_{\theta_{i}}^{T}\right)^{2}\sin^{2}\dfrac{\theta_{i}}{2}. \qquad (60)$$

Two conditions are considered for the joint connecting the $i^{th}$ body and its parent, and in each case, I will determine $\frac{\partial}{\partial \mathbf{q}} {}^{pr(i)}_{i}\mathbf{A}_4^T$. In the scenario where the joint is revolute, $\theta_i = f_i(\mathbf{q})$ is the joint variable, while $d_i$ remains constant. Referring to Definition 1, one can derive the following:

$$\frac{\partial}{\partial \mathbf{q}} {}^{pr(i)}_{i}\mathbf{A}_4^T = \begin{bmatrix} \frac{\partial}{\partial \mathbf{q}} {}^{pr(i)}_{i}\mathbf{A}^T & \mathbf{0}_{3n\times 1} \\ \mathbf{0}_{3\times n}^T & \mathbf{0}_{n\times 1} \end{bmatrix}. \tag{61}$$

---

[4] The derivations associated with Eq. 59 to Eq. 76 was previously published in Ehsani, H., M. Rostami and M. Parnianpour (2016). "A closed-form formula for the moment arm matrix of a general musculoskeletal model with considering joint constraint and motion rhythm." Multibody System Dynamics 36: 377-403.

Next, taking partial derivative with respect to the generalized coordinate vector from both sides of Eq. 60 yields

$$\frac{\partial}{\partial \mathbf{q}} {}^{pr(i)}_{i}\mathbf{A}^T = \mathbf{0}_{3n_q \times 3} + \frac{\partial}{\partial \mathbf{q}}\left(\tilde{\mathbf{V}}_{\theta_i}^T \sin f_i\right) + 2\frac{\partial}{\partial \mathbf{q}}\left(\left(\tilde{\mathbf{V}}_{\theta_i}^T\right)^2 \sin^2 \frac{f_i}{2}\right). \tag{62}$$

Applying Theorem 2 on Eq. 62 yields

$$\frac{\partial}{\partial \mathbf{q}}\left(\tilde{\mathbf{V}}_{\theta_i}^T \sin f_i\right) = \left(\tilde{\mathbf{V}}_{\theta_i}^T \cos f_i\right) \otimes \nabla f_i, \tag{63}$$

and,

$$\frac{\partial}{\partial \mathbf{q}}\left(\left(\tilde{\mathbf{V}}_{\theta_i}^T\right)^2 \sin^2 \frac{f_i}{2}\right) = \left(\left(\tilde{\mathbf{V}}_{\theta_i}^T\right)^2 \sin f_i\right) \otimes \nabla f_i. \tag{64}$$

Substituting Eq. 63, and Eq. 64 back into Eq. 62 reads

$$\frac{\partial}{\partial \mathbf{q}} {}^{pr(i)}_{i}\mathbf{A}^T = \left(\tilde{\mathbf{V}}_{\theta_i}^T \cos f_i + 2\left(\tilde{\mathbf{V}}_{\theta_i}^T\right)^2 \sin f_i\right) \otimes \nabla f_i. \tag{65}$$

As $\left(\tilde{\mathbf{V}}_{\theta_i}^T\right)^3 = \left(\left(\tilde{\mathbf{V}}_{\theta_i}\right)^3\right)^T = -\tilde{\mathbf{V}}_{\theta_i}^T$, the following equality holds:

$$\tilde{\mathbf{V}}_{\theta_i}^T {}^{pr(i)}_{i}\mathbf{A}^T = \tilde{\mathbf{V}}_{\theta_i}^T \cos f_i + 2\left(\tilde{\mathbf{V}}_{\theta_i}^T\right)^2 \sin f_i. \tag{66}$$

Comparing Eq. 65 and Eq. 66, the following conclusion is drawn:

$$\frac{\partial}{\partial \mathbf{q}} {}^{pr(i)}_{i}\mathbf{A}^T = \left( \tilde{\mathbf{V}}_{\theta_i}^T \, {}^{pr(i)}_{i}\mathbf{A}^T \right) \otimes \nabla f_i . \tag{67}$$

Substituting Eq. 67 into Eq. 61 and performing some manipulations yields

$$\frac{\partial}{\partial \mathbf{q}} {}^{pr(i)}_{i}\mathbf{A}_4^T = \left[ \begin{array}{c|c} \tilde{\mathbf{V}}_{\theta_i}^T \, {}^{pr(i)}_{i}\mathbf{A}^T & \mathbf{0}_{3\times 1} \\ \hline \mathbf{0}_{3\times 1}^T & 0 \end{array} \right] \otimes \nabla f_i . \tag{68}$$

This equation can be summarized as

$$\frac{\partial}{\partial \mathbf{q}} {}^{pr(i)}_{i}\mathbf{A}_4^T = \left( {}_{i}\hat{\mathbf{A}}_4^T \, {}^{pr(i)}_{i}\mathbf{A}_4^T \right) \otimes \nabla f_i \tag{69}$$

where ${}_{i}\hat{\mathbf{A}}_4$ is called a joint matrix and is defined as

$${}_{i}\hat{\mathbf{A}}_4^T = \left[ \begin{array}{c|c} \tilde{\mathbf{V}}_{\theta_i} & \mathbf{0}_{3\times 1} \\ \hline \mathbf{0}_{3\times 1}^T & 0 \end{array} \right] . \tag{70}$$

In case of a prismatic joint where $d_i = f_i(\mathbf{q})$, and is $\theta_i$ constant, Eq. 69 still holds. In this case, the joint matrix is derived as

$${}_{i}\hat{\mathbf{A}}_4 = \left[ \begin{array}{c|c} \mathbf{0}_{3\times 3} & {}^{i}_{pr(i)}\mathbf{A}\,\mathbf{V}_{d_i} \\ \hline \mathbf{0}_{3\times 1}^T & 0 \end{array} \right] . \tag{71}$$

Next, using the closed-from formula for $\frac{\partial}{\partial \mathbf{q}} {}^{pr(i)}_{i}\mathbf{A}_4^T$, I shall proceed with the derivation of the partial derivative of $\mathbf{B}_i$ with respect to the generalized coordinate vector.

As $\mathbf{B}_i = {}^{0}_{i}\mathbf{A}_4^T = {}^{pr(i)}_{i}\mathbf{A}_4^T \, {}^{0}_{pr(i)}\mathbf{A}_4^T$, employing Theorem 3 yields

$$\frac{\partial \mathbf{B}_i}{\partial \mathbf{q}} = \left( \frac{\partial}{\partial \mathbf{q}} {}^{pr(i)}_i \mathbf{A}_4^T \right) {}^{0}_{pr(i)} \mathbf{A}_4^T + \left( {}^{pr(i)}_i \mathbf{A}_4^T \otimes \mathbf{I}_{n_q} \right) \left( \frac{\partial \mathbf{B}_{pr(i)}}{\partial \mathbf{q}} \right). \tag{72}$$

Substituting Eq. 60 into this equation results in

$$\frac{\partial \mathbf{B}_i}{\partial \mathbf{q}} = \left[ \left( {}_i\hat{\mathbf{A}}_4^T \, {}^{pr(i)}_i \mathbf{A}_4^T \right) \otimes \nabla f_i \right] {}^{0}_{pr(i)} \mathbf{A}_4^T + \left( {}^{pr(i)}_i \mathbf{A}_4^T \otimes \mathbf{I}_{n_q} \right) \left( \frac{\partial \mathbf{B}_{pr(i)}}{\partial \mathbf{q}} \right) \tag{73}$$

With the application of the mixed-product property of the Kronecker multiplication, this equation is simplified as

$$\frac{\partial \mathbf{B}_i}{\partial \mathbf{q}} = \left( {}_i\hat{\mathbf{A}}_4^T \, {}^{0}_i \mathbf{A}_4^T \right) \otimes \nabla f_i + \left( {}^{pr(i)}_i \mathbf{A}_4^T \otimes \mathbf{I}_{n_q} \right) \left( \frac{\partial \mathbf{B}_{pr(i)}}{\partial \mathbf{q}} \right). \tag{74}$$

Eq. 74 delivers a recursive formulation to determine the first partial derivative of $\mathbf{B}_i$ with respect to the generalized coordinate. Expanding this equation would result in the following closed-from formulation:

$$\frac{\partial \mathbf{B}_i}{\partial \mathbf{q}} = \sum_{j \in Anc(i) \cup \{i\}} \mathbf{W}_{ji}^T \otimes \nabla f_j \tag{75}$$

where,

$$\mathbf{W}_{ji} = {}^{0}_j\mathbf{A}_4 \, {}_j\hat{\mathbf{A}}_4 \, {}^{j}_i\mathbf{A}_4 \,. \tag{76}$$

Using Eq. 76, the recursive formula of Eq. 74 can be restated as

$$\frac{\partial \mathbf{B}_i}{\partial \mathbf{q}} = \mathbf{W}_{ii}^T \otimes \nabla f_i + \left( {}^{pr(i)}_i \mathbf{A}_4^T \otimes \mathbf{I}_{n_q} \right) \left( \frac{\partial \mathbf{B}_{pr(i)}}{\partial \mathbf{q}} \right). \tag{77}$$

To determine $\frac{\partial^2 \mathbf{B}_i}{\partial \mathbf{q}^2}$, it is necessary to take the derivative with respect to the generalized coordinate vector from both sides of Equation 77, resulting in the following:

$$\frac{\partial^2 \mathbf{B}_i}{\partial \mathbf{q}^2} = \underbrace{\frac{\partial}{\partial \mathbf{q}}\left[\mathbf{W}_{ii}^T \otimes \nabla f_i\right]}_{1^{st}\ part} + \underbrace{\frac{\partial}{\partial \mathbf{q}}\left[\left(^{pr(i)}_{\phantom{i}i}\mathbf{A}_4^T \otimes \mathbf{I}_{n_q}\right)\left(\frac{\partial \mathbf{B}_{pr(i)}}{\partial \mathbf{q}}\right)\right]}_{2^{nd}\ part}. \tag{78}$$

Theorem 3 is used to simplify the 1$^{st}$ part in Eq. 78. Therefore,

$$\frac{\partial}{\partial \mathbf{q}}\left[\mathbf{W}_{ii}^T \otimes \nabla f_i\right] = \left(\mathbf{I}_4 \otimes \mathbf{U}_{n_q,n_q}\right)\left(\frac{\partial \mathbf{W}_{ii}^T}{\partial \mathbf{q}} \otimes \nabla f_i\right) + \mathbf{W}_{ii}^T \otimes \frac{\partial}{\partial \mathbf{q}}(\nabla f_i). \tag{79}$$

Given the definition of a gradient vector and what I stated for Definition 1, one can infer

$$\frac{\partial(\nabla f_i)}{\partial \mathbf{q}} = \text{vec}(\mathbf{H}_i) \tag{80}$$

where $\text{vec}(.)$ is an operator that flattens a 2D matrix into a 1D column vector (Neudecker 1969).

To evaluate $\frac{\partial \mathbf{W}_{ii}^T}{\partial \mathbf{q}}$, Eq. 76 and Theorem 1 are employed as follows:

$$\frac{\partial \mathbf{W}_{ii}^T}{\partial \mathbf{q}} = \left(_i\hat{\mathbf{A}}_4^T \otimes \mathbf{I}_{n_q}\right)\frac{\partial \mathbf{B}_i}{\partial \mathbf{q}}. \tag{81}$$

Of note, I have used the fact that $\frac{\partial_j\hat{\mathbf{A}}_4^T}{\partial \mathbf{q}}$ vanishes in deriving Eq. 81. Substituting Eq. 81 and Eq. 80 into Eq. 79, and performing certain manipulations yields

$$\frac{\partial}{\partial \mathbf{q}}\left[\mathbf{W}_{ii}^T \otimes \nabla f_i\right] = \left(_i\hat{\mathbf{A}}_4^T \otimes \mathbf{U}_{n_q,n_q}\right)\left(\frac{\partial \mathbf{B}_i}{\partial \mathbf{q}} \otimes \nabla f_i\right) \\ + \mathbf{W}_{ii}^T \otimes \text{vec}(\mathbf{H}_i). \tag{82}$$

To simplify the 2$^{nd}$ part of Eq. 78, at first, I apply Theorem 1 as follows:

$$\frac{\partial}{\partial \mathbf{q}}\left[\left({}^{pr(i)}_{\phantom{pr(i)}i}\mathbf{A}_4^T \otimes \mathbf{I}_{n_q}\right)\left(\frac{\partial \mathbf{B}_{pr(i)}}{\partial \mathbf{q}}\right)\right] = \frac{\partial\left({}^{pr(i)}_{\phantom{pr(i)}i}\mathbf{A}_4^T \otimes \mathbf{I}_{n_q}\right)}{\partial \mathbf{q}}\left(\frac{\partial \mathbf{B}_{pr(i)}}{\partial \mathbf{q}}\right) \\ + \left({}^{pr(i)}_{\phantom{pr(i)}i}\mathbf{A}_4^T \otimes \mathbf{I}_{n_q^2}\right)\left(\frac{\partial^2 \mathbf{B}_{pr(i)}}{\partial \mathbf{q}^2}\right) \quad (83)$$

On the other hand, by using Theorem 3 one can find

$$\frac{\partial\left({}^{pr(i)}_{\phantom{pr(i)}i}\mathbf{A}_4^T \otimes \mathbf{I}_{n_q}\right)}{\partial \mathbf{q}} = \left(\mathbf{I}_4 \otimes \mathbf{U}_{n_q,n_q}\right)\left[\frac{\partial\left({}^{pr(i)}_{\phantom{pr(i)}i}\mathbf{A}_4^T\right)}{\partial \mathbf{q}} \otimes \mathbf{I}_{n_q}\right]. \quad (84)$$

Substituting Eq. 69 into the right-hand side of this equation yields

$$\frac{\partial\left({}^{pr(i)}_{\phantom{pr(i)}i}\mathbf{A}_4^T \otimes \mathbf{I}_{n_q}\right)}{\partial \mathbf{q}} = \left(\mathbf{I}_4 \otimes \mathbf{U}_{n_q,n_q}\right)\left\{\left[\left({}_{i}\hat{\mathbf{A}}_4^T\, {}^{pr(i)}_{\phantom{pr(i)}i}\mathbf{A}_4^T\right) \otimes \nabla f_i\right] \otimes \mathbf{I}_{n_q}\right\}. \quad (85)$$

Using the associativity and mixed-product properties of the Kronecker multiplication, the right-hand side of Eq. 85 is simplified as

$$\frac{\partial\left({}^{pr(i)}_{\phantom{pr(i)}i}\mathbf{A}_4^T \otimes \mathbf{I}_{n_q}\right)}{\partial \mathbf{q}} = \left({}_{i}\hat{\mathbf{A}}_4^T\, {}^{pr(i)}_{\phantom{pr(i)}i}\mathbf{A}_4^T\right) \otimes \left(\mathbf{U}_{n_q,n_q}\left[\nabla f_i \otimes \mathbf{I}_{n_q}\right]\right). \quad (86)$$

Next, using the commutative rule of the Kronecker multiplication, one can deduce

$$\mathbf{U}_{n_q,n_q}\left[\nabla f_i \otimes \mathbf{I}_{n_q}\right] = \mathbf{I}_{n_q} \otimes \nabla f_i. \quad (87)$$

Therefore,

$$\frac{\partial\left({}^{pr(i)}_{\phantom{pr(i)}i}\mathbf{A}_4^T \otimes \mathbf{I}_{n_q}\right)}{\partial \mathbf{q}} = \mathbf{W}_{ii}^T \otimes \mathbf{I}_{n_q} \otimes \nabla f_i. \quad (88)$$

Consequently,

$$\frac{\partial^2 \mathbf{B}_i}{\partial \mathbf{q}^2} = \left( {}_i\hat{\mathbf{A}}_4^T \otimes \mathbf{U}_{n_q,n_q} \right)\left( \frac{\partial \mathbf{B}_i}{\partial \mathbf{q}} \otimes \nabla f_i \right) + \mathbf{W}_{ii}^T \otimes \text{vec}(\mathbf{H}_i)$$
$$+ \left[ \mathbf{W}_{ii}^T \otimes \mathbf{I}_{n_q} \otimes \nabla f_i \right]\left( \frac{\partial \mathbf{B}_{pr(i)}}{\partial \mathbf{q}} \right) \qquad (89)$$
$$+ \left( {}_i^{pr(i)}\mathbf{A}_4^T \otimes \mathbf{I}_{n_q^2} \right)\left( \frac{\partial^2 \mathbf{B}_{pr(i)}}{\partial \mathbf{q}^2} \right)$$

Eq. 89 provides a recursive formulation for obtaining the second partial derivative of $\mathbf{B}_i$ with respect to the generalized coordinate vector. However, if a closed-form formula for this quantity is desired, it is necessary to take partial derivatives from both sides of Eq. 75 with respect to the generalized coordinate vector. In the following explanation, I will demonstrate this derivation process.

Taking derivatives from both sides of Eq. 75 with respect to $\mathbf{q}$ yields

$$\frac{\partial^2 \mathbf{B}_i}{\partial \mathbf{q}^2} = \sum_{j \in Anc(i) \cup \{i\}} \left( \mathbf{I}_4 \otimes \mathbf{U}_{n_q \times n_q} \right)\left( \frac{\partial \mathbf{W}_{ji}^T}{\partial \mathbf{q}} \otimes \nabla f_j \right) + \mathbf{W}_{ji}^T \otimes \text{vec}(\mathbf{H}_i). \qquad (90)$$

To calculate the partial derivative of $\mathbf{W}_{ji}^T$ with respect to $\mathbf{q}$, I shall use Eq. 76 and Theorem 2 as follows:

$$\frac{\partial \mathbf{W}_{ji}^T}{\partial \mathbf{q}} = \frac{\partial \left( {}_i^j\mathbf{A}_4^T {}_j\hat{\mathbf{A}}_4^T \right)}{\partial \mathbf{q}} {}_j^0\mathbf{A}_4^T + \left[ \left( {}_i^j\mathbf{A}_4^T {}_j\hat{\mathbf{A}}_4^T \right) \otimes \mathbf{I}_{n_q} \right]\frac{\partial \mathbf{B}_j}{\partial \mathbf{q}}. \qquad (91)$$

Next, substituting Eq. 75 into Eq. 91 and using the fact that the partial derivative of the joint matrix with respect to the generalized coordinate vanishes, one may derive

$$\frac{\partial \mathbf{W}_{ji}^T}{\partial \mathbf{q}} = \frac{\partial {}_i^j\mathbf{A}_4^T}{\partial \mathbf{q}} {}_j\hat{\mathbf{A}}_4^T {}_j^0\mathbf{A}_4^T + \left[ \left( {}_i^j\mathbf{A}_4^T {}_j\hat{\mathbf{A}}_4^T \right) \otimes \mathbf{I}_{n_q} \right] \left[ \sum_{k \in Anc(j) \cup \{j\}} \mathbf{W}_{kj}^T \otimes \nabla f_k \right]. \qquad (92)$$

Applying the mixed-product property of the Kronecker multiplication on Eq. 92 results in

$$\frac{\partial \mathbf{W}_{ji}^T}{\partial \mathbf{q}} = \frac{\partial\, {}_i^j\mathbf{A}_4^T}{\partial \mathbf{q}}\, {}_j\hat{\mathbf{A}}_4^T\, {}_j^0\mathbf{A}_4^T + \sum_{k \in Anc(j) \cup \{j\}} \left( {}_i^j\mathbf{A}_4^T\, {}_j\hat{\mathbf{A}}_4^T \mathbf{W}_{kj}^T \right) \otimes \nabla f_k \qquad (93)$$

The only quantity that needs attention on the right-hand side of Eq. 93 is $\dfrac{\partial\, {}_i^j\mathbf{A}_4^T}{\partial \mathbf{q}}$. To determine this quantity, the same procedure used to determine $\dfrac{\partial \mathbf{B}_i}{\partial \mathbf{q}}$ shall be followed. In this case, the following closed-form formula is derived:

$$\frac{\partial\, {}_i^j\mathbf{A}_4^T}{\partial \mathbf{q}} = \sum_{s \in \Omega_{ij}} \left( {}_i^s\mathbf{A}_4^T\, {}_s\hat{\mathbf{A}}_4^T\, {}_s^j\mathbf{A}_4^T \right) \otimes \nabla f_s \qquad (94)$$

where $\Omega_{ij}$ is a set defined as

$$\Omega_{ij} = Anc(i) \cup \{i\} - Anc(j) - \{j\}. \qquad (95)$$

Substituting Eq. 94 back into Eq. 93 yields

$$\begin{aligned}\frac{\partial \mathbf{W}_{ji}^T}{\partial \mathbf{q}} &= \sum_{s \in \Omega_{ij}} \left( {}_i^s\mathbf{A}_4^T\, {}_s\hat{\mathbf{A}}_4^T\, {}_s^j\mathbf{A}_4^T\, {}_j\hat{\mathbf{A}}_4^T\, {}_j^0\mathbf{A}_4^T \right) \otimes \nabla f_s \\ &\quad + \sum_{k \in Anc(j) \cup \{j\}} \left( {}_i^j\mathbf{A}_4^T\, {}_j\hat{\mathbf{A}}_4^T \mathbf{W}_{kj}^T \right) \otimes \nabla f_k \end{aligned} \qquad (96a)$$

In other words,

$$\frac{\partial \mathbf{W}_{ji}^T}{\partial \mathbf{q}} = \sum_{s \in \Omega_{ij}} \left( {}_i^s\mathbf{A}_4^T\, {}_s\hat{\mathbf{A}}_4^T\, \mathbf{W}_{sj}^T \right) \otimes \nabla f_s + \sum_{k \in Anc(j) \cup \{j\}} \left( {}_i^j\mathbf{A}_4^T\, {}_j\hat{\mathbf{A}}_4^T \mathbf{W}_{kj}^T \right) \otimes \nabla f_k \qquad (96b)$$

The right-hand side of this equation can be reformulated as follows:

$$\frac{\partial \mathbf{W}_{ji}^T}{\partial \mathbf{q}} = \sum_{s \in Anc(j) \cup \{j\}} \overline{\mathbf{W}}_{sji}^T \otimes \nabla f_s \qquad (97)$$

where,

$$\bar{\mathbf{W}}_{sji} = \begin{cases} \mathbf{W}_{sj\ s}\hat{\mathbf{A}}_4{}_i^s\mathbf{A}_4 + \mathbf{W}_{kj\ j}\hat{\mathbf{A}}_4{}_i^j\mathbf{A}_4 & s \in \Omega_{ij} \\ \mathbf{W}_{kj\ j}\hat{\mathbf{A}}_4{}_i^j\mathbf{A}_4 & s \notin \Omega_{ij} \end{cases}. \tag{98}$$

Finally, by substituting Eq. 98 into Eq. 90, the following closed-form formula for the second derivative of $\mathbf{B}_i$ with respect to the generalized coordinate is derived:

$$\frac{\partial^2 \mathbf{B}_i}{\partial \mathbf{q}^2} = \sum_{j \in Anc(i) \cup \{i\}} \sum_{s \in Anc(i) \cup \{i\}} \left( \bar{\mathbf{W}}_{kji}^T \otimes \left[ \nabla f_j \otimes \nabla f_s \right] \right) + \sum_{j \in Anc(i) \cup \{i\}} \mathbf{W}_{ji}^T \otimes \text{vec}\left( \mathbf{H}_j \right). \tag{99}$$

### 3-3- The gradient vector and hessian matrix of joint variables

The gradient vector and Hessian matrix of the joint variables with respect to the generalized coordinate vector are vital components in the established formulation discussed in the previous section. These quantities must be available for each joint in order to utilize the derived formulations effectively. However, if the joint variables are constrained by a set of algebraic equations, finding explicit solutions for these nonlinear equations may not be feasible. In such cases, I propose three methods: Pre-modeling, numerical differentiation, and analytical projection.

In the pre-modeling approach, the generalized coordinates are swept within their acceptable range using a fixed step size. For each value of the generalized coordinate vector, the algebraic constraints are solved numerically using appropriate methods. The resulting discrete data points are then mathematically modeled using interpolation functions, such as natural cubic splines, due to the likelihood of a large number of data points. By differentiating these explicit smooth functions, the corresponding gradient vectors and Hessian matrices can be obtained. This method is particularly suitable for mechanisms with only one degree of freedom, but its applicability diminishes as the number of required generalized coordinates increases.

In contrast to the first approach, the second method employs numerical differentiation formulas. In this method, explicit equations for each joint variable with respect to the generalized coordinate vector are not required. Instead, at each instant of time, the algebraic constraint equations are solved numerically within the grids of a computational molecule. Fig. 2 illustrates this molecule for a mechanism with two degrees of freedom.

At a specific instant of time, denoted as $t^*$ in Fig. 2, the center of this computational molecule is constructed based on the current values of the generalized coordinates. The other nine abscissae of the molecule are obtained by taking small steps in the direction of each generalized coordinate, represented by $\Delta q_1$, and $\Delta q_2$ in Fig. 2. By solving the nonlinear algebraic equations at these ten abscissae, the values of the joint variables can be determined. Utilizing numerical differentiation formulas, the gradient vector and Hessian matrix corresponding to each joint variable can be calculated. Although pre-modeling is not required in this method, computational efficiency becomes a concern when the number of generalized coordinates and the grids of the computational molecules increase.

The third method presents an analytical approach for deriving the derivatives of the joint variables. For a multibody system that requires a minimum of $n_q$ generalized coordinates and has $m$ joint variables, denoted as $f_i$ for $i = 1, 2, \ldots, m$, the algebraic constraints can be represented as follows:

$$g_i(\mathbf{q}) = 0, \quad i = 1, 2, \ldots, m \tag{100}$$

where $g_i$ is a smooth function that is at least twice differentiable with respect to the generalized coordinate vector.

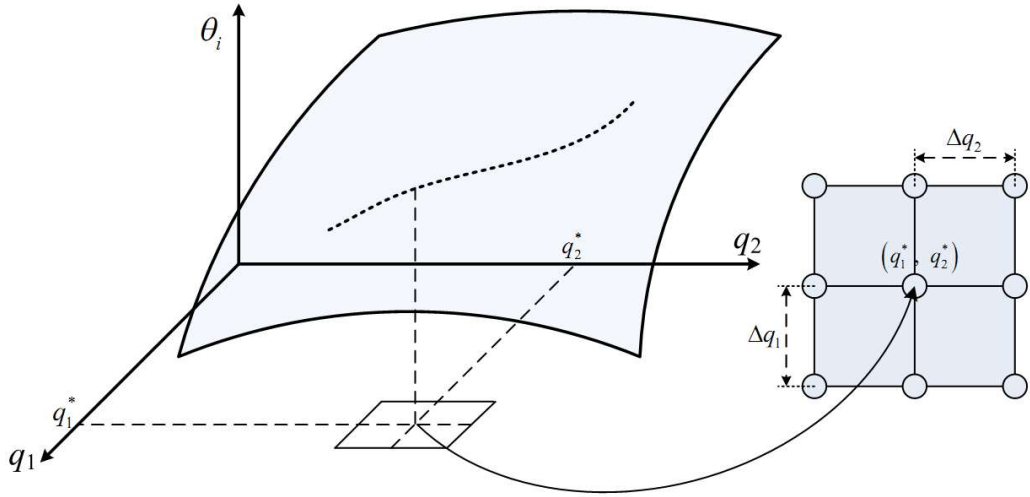

Figure 2 A 2D computational molecule to determine the gradient vector and hessian matrix of the $i^{th}$ joint

Taking derivative from both sides of Eq. 100 with respect to the vector $\mathbf{q}$ yields

$$\sum_{j=1}^{m} \frac{\partial g_i}{\partial f_j} \nabla f_j = -\frac{\partial g_i}{\partial \mathbf{q}}, \quad i = 1, 2, \ldots, m. \tag{101}$$

This equation is restated as follows via the Kronecker multiplication:

$$\sum_{j=1}^{m} \overline{\mathbf{g}}_j \otimes \nabla f_j = -\frac{\partial \mathbf{G}}{\partial \mathbf{q}} \tag{102}$$

where,

$$\mathbf{G} = \{g_1 \quad g_2 \quad \cdots \quad g_m\}^T \tag{103}$$

and,

$$\overline{\mathbf{g}}_j = \left\{ \frac{\partial g_1}{\partial f_j} \quad \frac{\partial g_2}{\partial f_j} \quad \cdots \quad \frac{\partial g_m}{\partial f_j} \right\}^T. \tag{104}$$

To solve the system of equations shown in Eq. 102, I shall use an elimination method.

Let

$$\boldsymbol{\eta}^{(m)} = \sum_{j=1}^{m} \overline{\mathbf{g}}_j \otimes \nabla f_j. \tag{105}$$

Therefore,

$$\boldsymbol{\eta}^{(m)} = \boldsymbol{\eta}^{(m-1)} + \overline{\mathbf{g}}_m \otimes \nabla f_m. \tag{106}$$

Substituting this equation into Eq. 102 results in

$$\boldsymbol{\eta}^{(m-1)} + \overline{\mathbf{g}}_m \otimes \nabla f_m = -\frac{\partial \mathbf{G}}{\partial \mathbf{q}}. \tag{107}$$

Pre-multiplying both sides of Eq. 107 by $\left( \overline{\mathbf{g}}_m^T \otimes \mathbf{I}_{n_q} \right)$ and performing certain manipulations yield

$$\nabla f_m = -\frac{1}{\|\overline{\mathbf{g}}_m\|^2} \left( \overline{\mathbf{g}}_m^T \otimes \mathbf{I}_{n_q} \right) \left( \boldsymbol{\eta}^{(m-1)} + \frac{\partial \mathbf{G}}{\partial \mathbf{q}} \right) \tag{108}$$

where $\|\cdot\|$ shows a 2-norm for vectors.

By back substituting Eq. 108 into Eq. 102, one can find

$$\left( \boldsymbol{\gamma}_m \otimes \mathbf{I}_{n_q} \right) \boldsymbol{\eta}^{(m-1)} = -\left( \boldsymbol{\gamma}_m \otimes \mathbf{I}_{n_q} \right) \frac{\partial \mathbf{G}}{\partial \mathbf{q}} \tag{109}$$

where

$$\boldsymbol{\gamma}_m = \mathbf{I}_m - \frac{\overline{\mathbf{g}}_m \overline{\mathbf{g}}_m^T}{\|\overline{\mathbf{g}}_m\|^2}. \tag{110}$$

Next, using the recursive formula given in Eq. 106, I expand Eq. 109 as follows:

$$\left(\boldsymbol{\gamma}_m \otimes \mathbf{I}_{n_q}\right)\left(\boldsymbol{\eta}^{(m-2)} + \overline{\mathbf{g}}_{m-1} \otimes \nabla f_{m-1}\right) = -\left(\boldsymbol{\gamma}_m \otimes \mathbf{I}_{n_q}\right)\frac{\partial \mathbf{G}}{\partial \mathbf{q}} \tag{111}$$

Similar to what was done for Eq. 107, the following is derived from Eq. 111:

$$\nabla f_{m-1} = -\frac{1}{\|\boldsymbol{\gamma}_m \overline{\mathbf{g}}_{m-1}\|^2}\left(\overline{\mathbf{g}}_{m-1}^T \boldsymbol{\gamma}_m \otimes \mathbf{I}_{n_q}\right)\left(\boldsymbol{\eta}^{(m-2)} + \frac{\partial \mathbf{G}}{\partial \mathbf{q}}\right). \tag{112}$$

Back substituting this equation into Eq. 111 and performing the necessary manipulations result in

$$\left(\boldsymbol{\gamma}_{m-1} \otimes \mathbf{I}_{n_q}\right)\boldsymbol{\eta}^{(m-2)} = -\left(\boldsymbol{\gamma}_{m-1} \otimes \mathbf{I}_{n_q}\right)\frac{\partial \mathbf{G}}{\partial \mathbf{q}} \tag{113}$$

where

$$\boldsymbol{\gamma}_{m-1} = \boldsymbol{\gamma}_m \boldsymbol{\sigma}_{m-1} \tag{114}$$

and,

$$\boldsymbol{\sigma}_{m-1} = \mathbf{I}_m - \frac{1}{\|\boldsymbol{\gamma}_m \overline{\mathbf{g}}_{m-1}\|^2} \overline{\mathbf{g}}_{m-1} \overline{\mathbf{g}}_{m-1}^T \boldsymbol{\gamma}_m. \tag{115}$$

This procedure can be continued in a similar fashion. To this end, the following recursive formulations are used

$$\boldsymbol{\sigma}_i = \mathbf{I}_m - \frac{1}{\|\boldsymbol{\gamma}_{i+1} \overline{\mathbf{g}}_i\|^2} \overline{\mathbf{g}}_i \overline{\mathbf{g}}_i^T \boldsymbol{\gamma}_{i+1}, \quad i = m, m-1, ..., 2 \tag{116}$$

and

$$\gamma_i = \gamma_{i+1}\sigma_i, \quad i = m, m-1, \ldots, 2. \tag{117}$$

Of note, I have assumed that $\gamma_{m+1} = \mathbf{I}_m$.

Using Eq. 116 and Eq. 117 in a recursive fashion, one can use the following formulation to find $\nabla f_i$ for $i = 1, 2, \ldots, m$

$$\nabla f_i = -\frac{1}{\|\gamma_{i+1}\bar{\mathbf{g}}_i\|^2}\left(\bar{\mathbf{g}}_i^T \gamma_{i+1} \otimes \mathbf{I}_{n_q}\right)\left(\boldsymbol{\eta}^{(i-1)} + \frac{\partial \mathbf{G}}{\partial \mathbf{q}}\right), \quad i = 1, 2, \ldots, m \tag{118}$$

where,

$$\begin{aligned}\boldsymbol{\eta}^{(0)} &= \mathbf{0}_{(m \times n_q) \times 1} \\ \boldsymbol{\eta}^{(i)} &= \boldsymbol{\eta}^{(i-1)} + \bar{\mathbf{g}}_i \otimes \nabla f_i, \quad i = 1, 2, \ldots, m\end{aligned} \tag{119}$$

Taking derivatives from both sides of Eq. 102 with respect to the generalized coordinate vector (Theorem 3 must be employed to do so) and performing necessary manipulations result in

$$\sum_{j=1}^{m} \bar{\mathbf{g}}_j \otimes \text{vec}(\mathbf{H}_j) = \Gamma \tag{120}$$

where $\Gamma$ is a vector function with dimensions consistent with the problem. Since the structure of Eq. 120 resembles Eq. 102, a similar algorithm is employed to determine $\text{vec}(\mathbf{H}_j)$. Given that projection matrices are used for eliminations in this method, it shall be referred to as the analytical projection method.

### 4-3- Simulations

In order to validate the effectiveness of the developed algorithms, this section will simulate two constrained mechanisms. The kinematic analysis of these mechanisms will be conducted using both the developed algorithm and the method proposed in (Shabana 2005) for verification purposes.

## A) Four-bar planar closed chain linkage

Figure 3 illustrates a schematic of this mechanism. It consists of three moving bodies and the ground interconnected by revolute joints. As there is a closed kinematic chain formed between the moving bodies and the ground, only one generalized coordinate is needed to fully determine the kinematics of this planar mechanism. Let the CoM height of the second body be denoted as the generalized coordinate. Referring to Fig. 3, the algebraic constraints between the joint variables ($\theta_1$, $\theta_2, \theta_3$) and the generalized coordinate ($q$) can be expressed as follows:

$$\begin{cases} l_1 S_1 + l_{cm2} S_{12} - q = 0 \\ (l_2 - l_{cm2}) S_{12} + l_3 S_{123} + q = 0 \\ l_1 C_1 + l_2 C_{12} + l_3 C_{123} - l_0 = 0 \end{cases} \tag{121}$$

where,

$$\begin{aligned} S_{12...m} &= \sin\left(\sum_{i=1}^{m} \theta_i\right) \\ C_{12...m} &= \cos\left(\sum_{i=1}^{m} \theta_i\right) \end{aligned} \tag{122}$$

The parameters of Eq. 121 and Eq. 122 are defined in Fig. 3. Also, the numerical values of these parameters are given in Table 1.

A desired trajectory for the generalized coordinate of this mechanism is generated using piecewise Hermite polynomials, as depicted in Fig. 4. Furthermore, it is expected to determine the velocity and acceleration of point p as the generalized coordinate changes according to this desired trajectory.

Table 1 Numerical values of the parameters of the four-bar linkage

| parameter | $l_0$ | $l_1$ | $l_2$ | $l_3$ | $l_{cm2}$ | $l_p$ |
|---|---|---|---|---|---|---|
| value (units) | 5 | 2 | 4 | $\sqrt{5}$ | 2 | $\dfrac{\sqrt{5}}{2}$ |

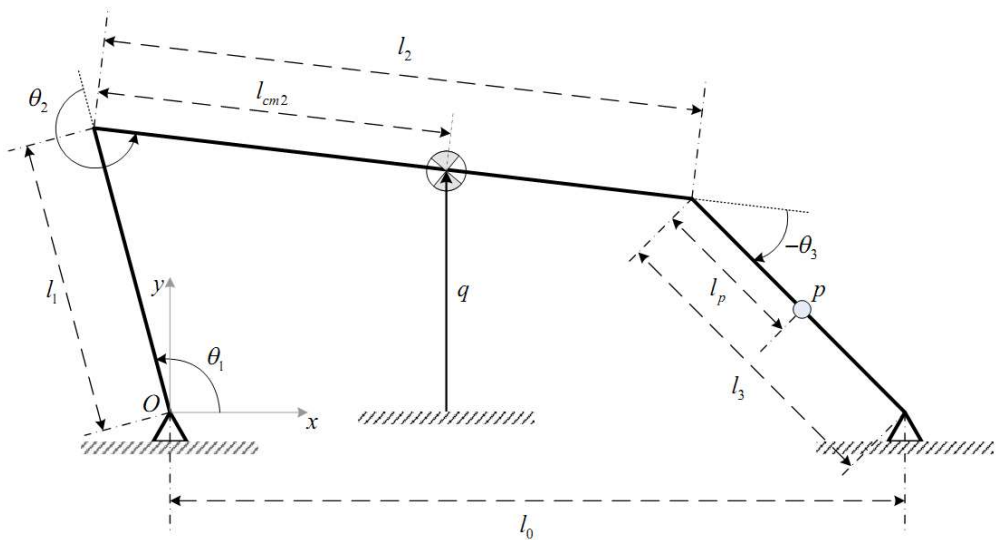

Figure 3 A schematic of the planar four-bar linkage used in this section. The inertial reference frame is located at point O.

To determine the first and second derivatives of the joint variables with respect to the generalized coordinate, the pre-modeling approach is utilized. With the known parameter values of this mechanism, the range of motion for the generalized coordinate spans from -2 to 2 units. Within this interval, the coordinate is swept with a step length of 0.01, and the algebraic constraints are solved for each value to determine the joint variables. Subsequently, the results are expressed as smooth functions of the generalized coordinate using cubic splines. These results are depicted in Fig. 5. By differentiating these smooth functions, the first and second derivatives of the joint variables with respect to the generalized coordinate are obtained.

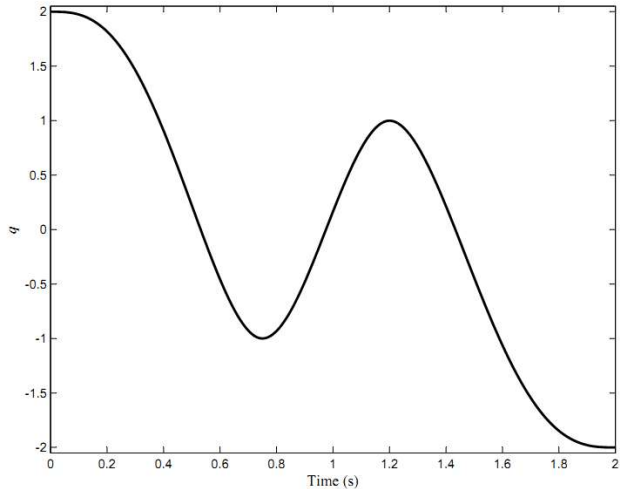

Figure 4 The desired trajectory of the generalized coordinate of the four-bar linkage mechanism

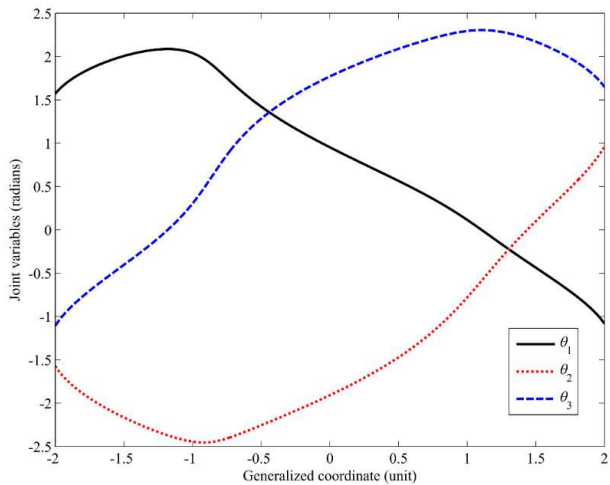

Figure 5 Joint variables of the four-bar linkage expressed as smooth functions of the genderized coordinate of the mechanism.

Using the first and second derivatives of the joint variables with respect to the generalized coordinate and the algorithm presented in this section, the velocity and acceleration of point p is calculated. To verify the results, this problem is also solved using a traditional method of kinematic analysis. The results obtained from these two methods were equal to 7 significant figures. The magnitudes of the velocity and acceleration vectors are shown in Fig. 6.

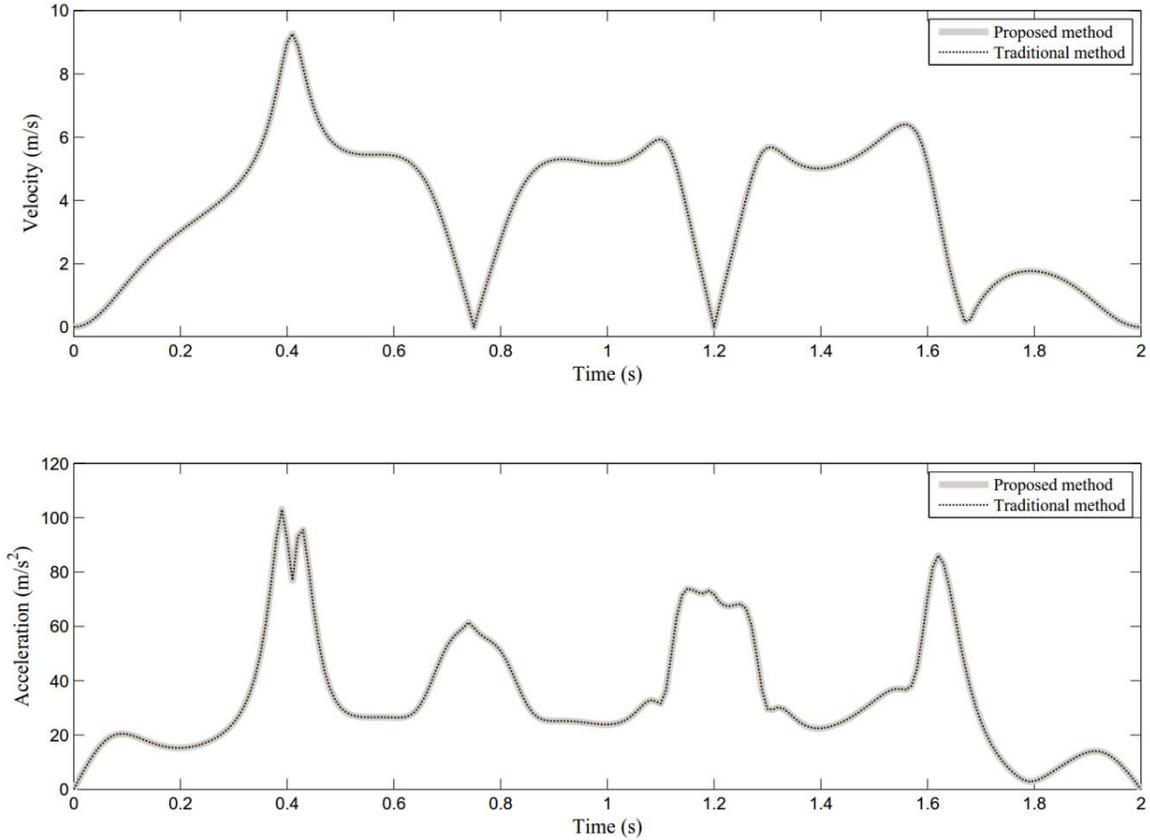

Figure 6: Magnitude of velocity and acceleration of point *p* in the four-bar linkage mechanism shown in Fig. 3 while the generalized coordinate follows the desired trajectory shown in Fig. 4. To validate the results obtained from the proposed method in this dissertation, results obtained from a traditional method of kinematic analysis are also shown.

**B) Five-bar planar closed chain linkage**

In this subsection, the 2-dof mechanism shown in Fig. 6 will be examined. In this mechanism, the joint variables of the first and second joints will be considered as the generalized coordinates. Due to the existence of a loop in this mechanism, the following set of constraints exist between the joint variables and the generalized coordinates:

$$\begin{cases} l_1\,S_1+l_2\,S_{12}+l_3\,S_{123}+l_4\,S_{1234}=0 \\ l_1\,C_1+l_2\,C_{12}+l_3\,C_{123}+l_4\,C_{1234}=l_0 \end{cases} \tag{123}$$

where $S_{12...m}$ and $C_{12...m}$ are defined in Eq. 122. Numerical values of the parameters of this mechanism are given in Table 2.

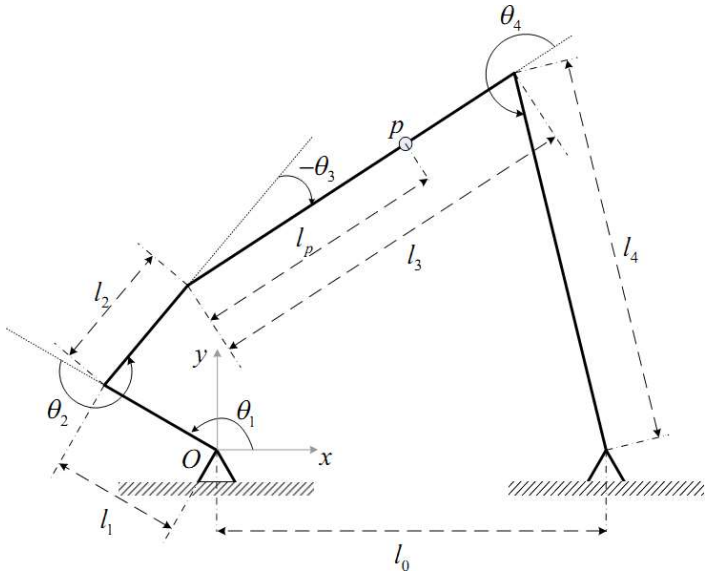

Figure 7 A schematic of the planar five-bar linkage used in this

Table 2 Numerical values of the parameters of the five-bar linkage

| parameter | $l_0$ | $l_1$ | $l_2$ | $l_3$ | $l_4$ | $l_p$ |
|---|---|---|---|---|---|---|
| value (units) | 3 | 1 | 1 | 3 | 3 | 2 |

Similar to the previous example, desired trajectories for the generalized coordinates of this mechanism are generated using piecewise Hermite polynomials. These trajectories are depicted in Fig. 7. The objective is to determine the velocity and acceleration of point *p* as the generalized coordinates follow these trajectories.

To determine the first and second derivatives of the joint variables with respect to the generalized coordinates, I shall use the analytical projection method. Similar to the previous mechanism, the results are also obtained using the method established in (Shabana 2005). The results, illustrated in Fig. 9, were identical to each other up to 6 significant figures.

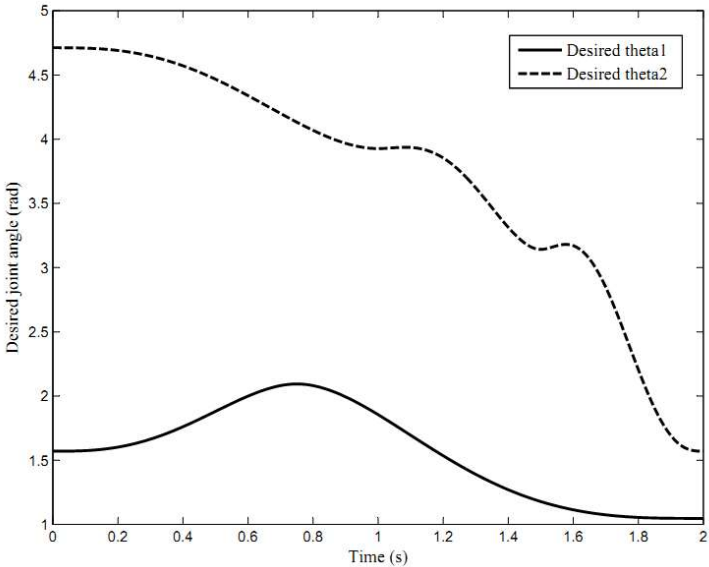

Figure 8 The desired trajectories of the generalized coordinates of the five-bar linkage mechanism

Although the projection method is employed for this mechanism, it is worth noting that the numerical differentiation method would yield the same results. To demonstrate this, in Fig. 10, the magnitude of the gradient vector for the third and fourth joints of this mechanism is compared with those obtained from the projection method. These results are accurate up to 7 significant figures.

While both the projection and numerical differentiation methods produce identical outcomes, it is important to consider the step length required for the latter. When there is a highly nonlinear and volatile relationship between the joint variable and the generalized coordinates, the fidelity of the results obtained through the numerical differentiation method may become questionable if the step length is not small enough. To illustrate this, Fig. 11 showcases the plot of the fourth joint manifold of the five-bar mechanism against the generalized coordinates. Due to the highly nonlinear behavior of this manifold, selecting a sufficiently small step length becomes crucial to accurately capture its variations with respect to the generalized coordinates. As a result, in such cases, the numerical differentiation method becomes computationally inefficient and impractical.

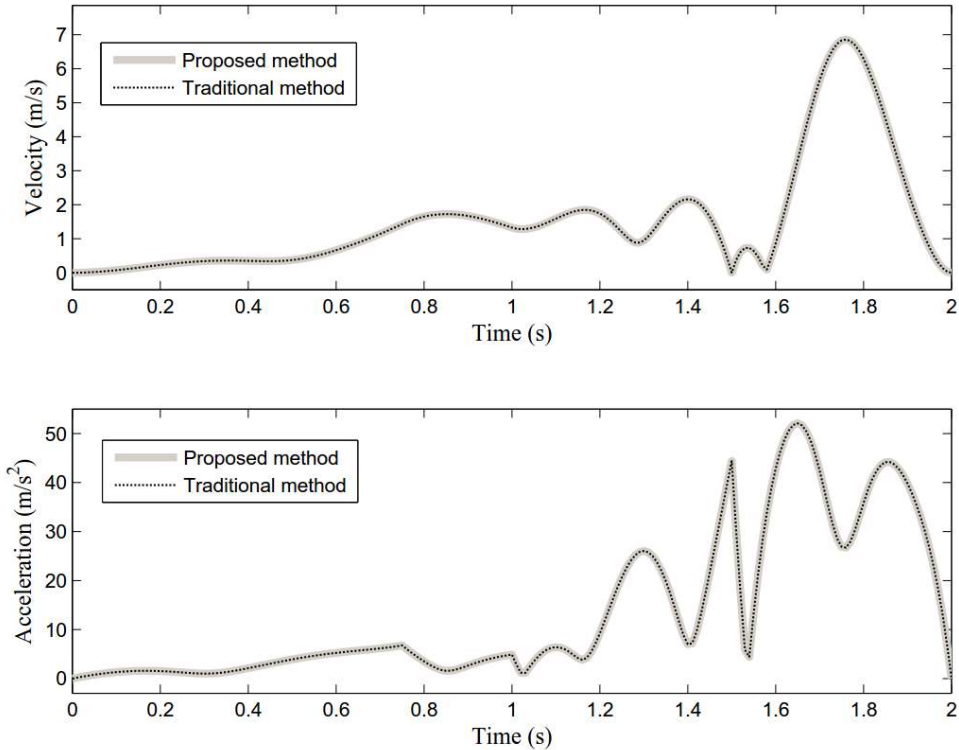

Figure 9. The magnitude of velocity and acceleration vector of point *p* of the mechanism depicted in Fig. 7 versus time. Thick solid lines show the outcomes of the closed-form formulae developed in this study, while the dotted lines have been obtained from the traditional method.

## 4- Dynamics

In this section, the derivation of the governing equations for the musculoskeletal system of the human body will be presented utilizing the virtual work principle. The virtual work for an arbitrary body $i$ can be divided into two components: the virtual work associated with inertial forces, i.e. $\delta W_I^i$, and the virtual work related to external effects, i.e., $\delta W_E^i$. Initially, my focus will be on examining the inertial forces and subsequently, I will proceed with discussing the virtual work attributed to external effects. Finally, I will extend this discussion to encompass a multibody system and derive the governing equations for the entire system.

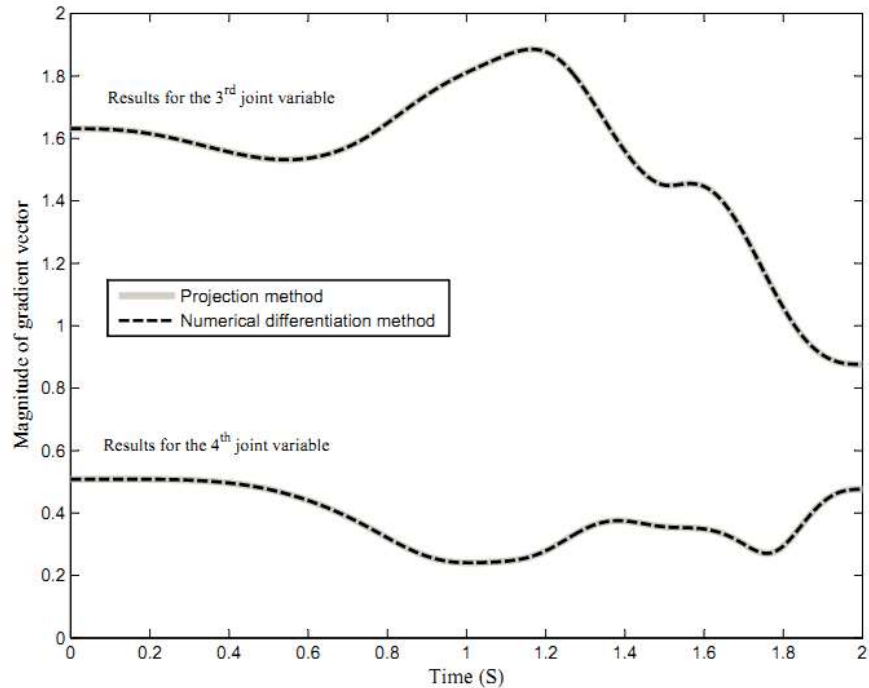

Figure10 The magnitude of the gradient vector for the third and fourth joint of the mechanism described in Fig. 7, with respect to time. These vectors have been computed from the projection method (thick solid lines) and numerical differentiation approach (dashed lines).

### 1-4- Inertial virtual work for an arbitrary body

Consider a point mass of $dm_i$ situated at an arbitrary point $p$ on the $i^{th}$ body (as shown in Fig. 8). The virtual work associated with the inertial forces for this point mass can be readily obtained. By integrating this virtual work over the entire body, one can deduce the following expression for the inertial virtual work of the $i^{th}$ body:

$$\delta W_I^i = \int_{R_i} -trace\left[\left\{-\dfrac{^0\mathbf{a}_p^i}{0}\right\}\left\{\left(\delta{^0\mathbf{r}_p^i}\right)^T \mid 0\right\}\right]dm_i \tag{124}$$

where $trace[.]$ shows the trace operator for a matrix.

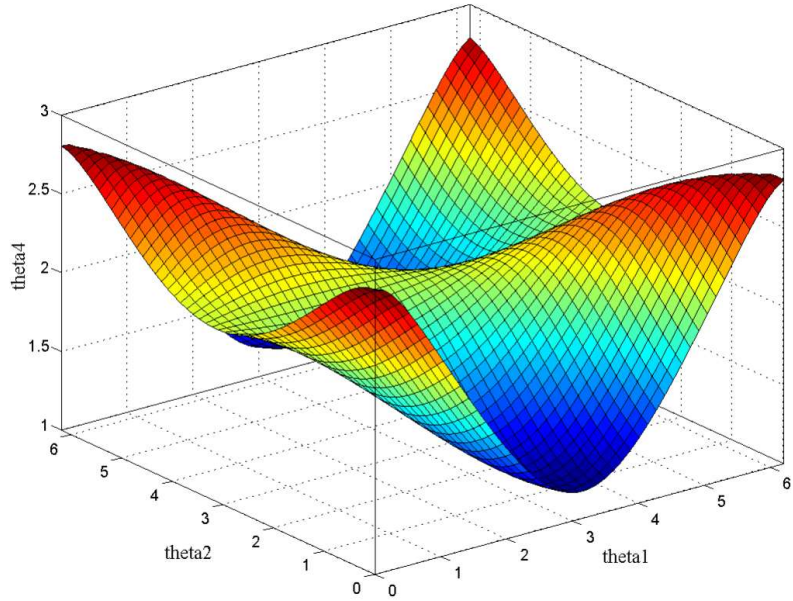

Figure 11 The manifold of the 4$^{th}$ joint with respect to the generalized coordinates

Before proceeding with the calculation of this integral, according to what was derived in the previous section, it is important to note that the following relationship holds:

$$^0\mathbf{a}_p^i\left(\delta{^0\mathbf{r}_p^i}\right)^T = \underbrace{\mathbf{D}_i^T\left(\left\{\dfrac{^i\mathbf{r}_p}{1}\right\}\otimes\ddot{\mathbf{q}}\right)\left(\left\{{^i\mathbf{r}_p^T} \mid 1\right\}\otimes\delta\mathbf{q}^T\right)\mathbf{D}_i}_{1^{st}\ part}$$
$$+\underbrace{(\mathbf{D}_i')^T\left(\left\{\dfrac{^i\mathbf{r}_p}{1}\right\}\otimes\dot{\mathbf{q}}^{[2]}\right)\left(\left\{{^i\mathbf{r}_p^T} \mid 1\right\}\otimes\delta\mathbf{q}^T\right)\mathbf{D}_i}_{2^{nd}\ part} \tag{125}$$

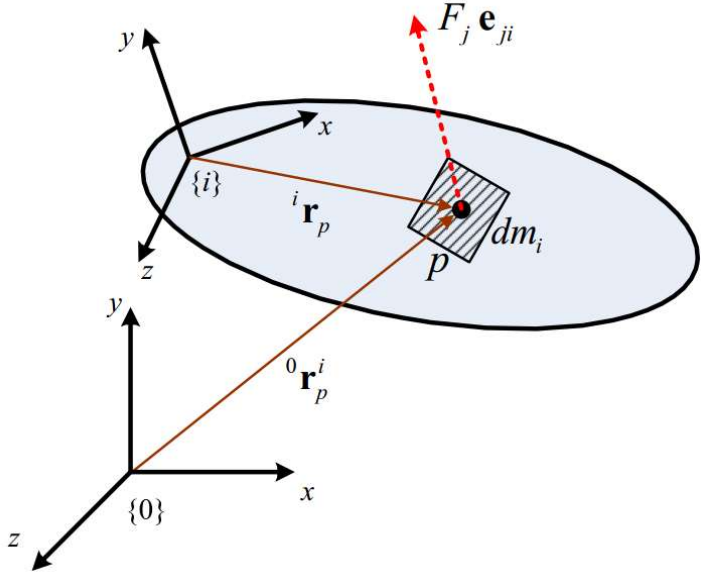

Figure 12: The schematic of the $i^{th}$ body depicts a differential mass element positioned at point $p$ within the body. Additionally, an external force is applied to this point, represented by the magnitude $F_j$ and the direction $\mathbf{e}_{ij}$.

Using the mixed-product property, this equation is rewritten as:

$$^0\mathbf{a}_p^i\left(\delta\,^0\mathbf{r}_p^i\right)^T = \underbrace{\mathbf{D}_i^T\left[\bar{\mathbf{J}}_i \otimes \left(\ddot{\mathbf{q}}\delta\mathbf{q}^T\right)\right]\mathbf{D}_i}_{1^{st}\ \text{part}} \\ + \underbrace{\left(\mathbf{D}_i'\right)^T\left[\bar{\mathbf{J}}_i \otimes \left(\dot{\mathbf{q}}^{[2]}\delta\mathbf{q}^T\right)\right]\mathbf{D}_i}_{2^{nd}\ \text{part}} \qquad (126)$$

where

$$\bar{\mathbf{J}}_i = \left[\begin{array}{c|c} {^i\mathbf{r}_p}\,{^i\mathbf{r}_p^T} & {^i\mathbf{r}_p} \\ \hline {^i\mathbf{r}_p^T} & 1 \end{array}\right] \qquad (127)$$

Accordingly, the inertial virtual work reads

$$\delta W_I^i = -\int_{R_i} trace\left[1^{st}\ \text{part}\right]dm_i - \int_{R_i} trace\left[2^{nd}\ \text{part}\right]dm_i. \qquad (128)$$

To calculate the above equation, I shall begin with the first part. In doing so, I need to consider two points. Firstly, since trace and integral are linear operators, one can state

$$\int_{R_i} trace\left[1^{st}\ part\right] dm_i = trace\left[\int_{R_i} \left(1^{st}\ part\right) dm_i\right]. \tag{129}$$

Furthermore,

$$\int_{R_i} \bar{\mathbf{J}}_i\, dm_i = \mathbf{J}_i \tag{130}$$

and,

$$\mathbf{J}_i = \left[\begin{array}{c|c} \mathbf{I}_p^i & m_i{}^i\mathbf{r}_{cm} \\ \hline m_i{}^i\mathbf{r}_{cm}^T & m_i \end{array}\right] \tag{131}$$

where $\mathbf{I}_p^i$ is the moment of inertia tensor of the $i^{th}$ body calculated at point $p$. Consequently,

$$\int_{R_i} \left(1^{st}\ part\right) dm_i = \mathbf{D}_i^T \left[\mathbf{J}_i \otimes \left(\ddot{\mathbf{q}}\delta\mathbf{q}^T\right)\right] \mathbf{D}_i. \tag{132}$$

Using the above equation and $\ddot{\mathbf{q}}\delta\mathbf{q}^T = \delta\mathbf{q}\ddot{\mathbf{q}}^T$ result in

$$\int_{R_i} trace\left[1^{st}\ part\right] dm_i = trace\left[\mathbf{D}_i^T \left[\mathbf{J}_i \otimes \left(\delta\mathbf{q}\ddot{\mathbf{q}}^T\right)\right] \mathbf{D}_i\right]. \tag{133}$$

Given the relationship between vec(.) and *trace*[.] operators (Neudecker 1969), I shall rewrite the right-hand side of this equation as follows:

$$trace\left[\mathbf{D}_i^T \left[\mathbf{J}_i \otimes \left(\ddot{\mathbf{q}}\delta\mathbf{q}^T\right)\right] \mathbf{D}_i\right] = \left(\text{vec}(\mathbf{D}_i)\right)^T \left(\mathbf{I}_4 \otimes \mathbf{J}_i \otimes \left(\delta\mathbf{q}\ddot{\mathbf{q}}^T\right)\right) \text{vec}(\mathbf{D}_i) \tag{134}$$

Using the mixed-product and associativity properties of the Kronecker multiplication, the right-hand side of this equation is factorized as follows:

$$\left(\text{vec}(\mathbf{D}_i)\right)^T \left(\mathbf{I}_4 \otimes \mathbf{J}_i \otimes \left(\delta\mathbf{q}\ddot{\mathbf{q}}^T\right)\right) \text{vec}(\mathbf{D}_i) = \xi_i \zeta_i \tag{135}$$

where,

$$\xi_i = \left(\text{vec}(\mathbf{D}_i)\right)^T \left(\mathbf{I}_4 \otimes \mathbf{J}_i \otimes \delta\mathbf{q}\right) \tag{136}$$

and,

$$\zeta_i = \left(\mathbf{I}_{16} \otimes \ddot{\mathbf{q}}^T\right)\text{vec}(\mathbf{D}_i). \tag{137}$$

In what follows, I will show how to separate $\delta\mathbf{q}$ and $\ddot{\mathbf{q}}$ from the other variables in Eq. 136 and Eq. 137, respectively.

Eq, 137 is restated as follows:

$$\zeta_i = \left(\mathbf{I}_4 \otimes \left[\mathbf{I}_4 \otimes \ddot{\mathbf{q}}^T\right]\right)\text{vec}(\mathbf{D}_i). \tag{138}$$

In other words,

$$\zeta_i = \text{vec}\left(\left[\mathbf{I}_4 \otimes \ddot{\mathbf{q}}^T\right]\mathbf{D}_i\right). \tag{139}$$

Therefore, given the properties of the vec(.) operator (Neudecker 1969), I shall rewrite this formula as follows:

$$\zeta_i = \left(\mathbf{D}_i^T \otimes \mathbf{I}_4\right)\text{vec}\left(\mathbf{I}_4 \otimes \ddot{\mathbf{q}}^T\right). \tag{140}$$

Using the relationship between the Kronecker product and the vec(.) operator (Magnus and Neudecker 1979), this equation is restated as

$$\zeta_i = \left(\mathbf{D}_i^T \otimes \mathbf{I}_4\right)\left(\mathbf{I}_4 \otimes \mathbf{U}_{n_q,4}\right)\mathbf{C}_1\ddot{\mathbf{q}} \tag{141}$$

where,

$$\mathbf{C}_1 = \text{vec}(\mathbf{I}_4) \otimes \mathbf{I}_{n_q}. \tag{142}$$

The steps to simplify $\xi_i$ are essentially the same. Consequently, one would obtain the following:

$$\xi_i = \delta \mathbf{q}^T \mathbf{C}_1^T \left( \mathbf{J}_i \otimes \mathbf{U}_{n_q,4}^T \right) \left( \mathbf{D}_i \otimes \mathbf{I}_4 \right). \tag{143}$$

Substituting these equations into Eq. 140.3 yields

$$\int_{R_i} trace\left[1^{st} \text{ part}\right] dm_i = \delta \mathbf{q}^T \mathbf{M}_i \ddot{\mathbf{q}} \tag{144}$$

where,

$$\mathbf{M}_i = \mathbf{C}_1^T \left( \mathbf{J}_i \otimes \mathbf{U}_{n_q,4}^T \right) \left( \mathbf{D}_i \mathbf{D}_i^T \otimes \mathbf{I}_4 \right) \left( \mathbf{I}_4 \otimes \mathbf{U}_{n_q,4} \right) \mathbf{C}_1 \tag{145}$$

This equation illustrates the contribution of the $i^{th}$ body to the mass matrix of the entire multibody system. Further elaboration on this concept will be provided later on.

Using a similar approach, the second part of the integral, i.e., $\int_{R_i} trace\left[2^{nd} \text{ part}\right] dm_i$, is determined as follows:

$$\int_{R_i} trace\left[2^{nd} \text{ part}\right] dm_i = \delta \mathbf{q}^T \mathbf{N}_i \dot{\mathbf{q}}. \tag{146}$$

where $\mathbf{N}_i$ contains the Coriolis and centrifugal effects associated with the $i^{th}$ body and is calculated as follows:

$$\mathbf{N}_i = \mathbf{C}_1^T \left( \mathbf{J}_i \otimes \mathbf{U}_{n_q,4}^T \right) \left( \mathbf{D}_i \mathbf{D}_i^T \otimes \mathbf{I}_4 \right) \left( \mathbf{I}_4 \otimes \mathbf{U}_{n_q,4} \right) \mathbf{C}_1. \tag{147}$$

Using these results, the inertial virtual work corresponding to the $i^{th}$ body is determined as follows:

$$\delta W_I^i = -\delta \mathbf{q}^T \left( \mathbf{M}_i \ddot{\mathbf{q}} + \mathbf{N}_i \dot{\mathbf{q}} \right). \tag{148}$$

## 2-4-Virtual work of the external effects

To calculate the virtual work associated with the external effects, two types of forces are considered: forces due to the musculotendon units and the weight force. Each case will be discussed separately.

### 1-2-4 Virtual work corresponding to the forces of the musculotendon units[5]

Assuming that there are $n_{mt}$ musculotendon units present in the model, the force of the $j^{th}$ unit is represented by $F_{mt}^{j}$. The tuple $\mathbf{F}_{mt}$ consists of all the forces of the musculotendon units. In other words

$$\mathbf{F}_{mt} = \left\{ F_{mt}^{1} \quad F_{mt}^{2} \quad \cdots \quad F_{mt}^{n_{mt}} \right\}^{T}. \tag{149}$$

Since an arbitrary muscle does not exert force on all the bodies of a multibody system, to determine the effect of a muscle on the rigid bodies of the model, I shall define the "Attachment sequence".

### Definition 2: Attachment sequence for the kth musculotendon unit

The attachment sequence $\Gamma_k$ is a sequence that includes all the identification numbers of the bodies on which the $k^{th}$ musculotendon unit acts. In this sequence, the "Ground" body is identified as "0", and it may be included as a member. Mathematically, the attachment sequence is defined as follows:

---

[5] The materials of this subsection were previously published in Ehsani, H., M. Rostami and M. Parnianpour (2016). "A closed-form formula for the moment arm matrix of a general musculoskeletal model with considering joint constraint and motion rhythm." Multibody System Dynamics **36**: 377-403.

$$T \mapsto \{0,1,2,\ldots,n_b\}$$
$$j \mapsto \Gamma_k(j)$$

where $T$ is an inductive subset of the natural numbers, i.e., N.

The term "sequence" implies that the order of the indices in the attachment sequence matters, unlike in a set. It is important to note that the first member of this attachment sequence corresponds to the body of origin, while the last member corresponds to the body of insertion. For the sake of uniformity, I have assumed that the body of origin has the smaller identification number. In other words, the body of origin is the one located closer to the ground in comparison to the body of insertion.

Using a conventional definition for a muscle, considering only two bodies (the origin and insertion bodies) should suffice to define this sequence; however, this is not the case when geometric objects are used to establish the path of a musculotendon unit. When a musculotendon unit passes through or wraps around the objects, the bodies are influenced and should be therefore included in the attachment sequence. For instance, for the musculotendon unit depicted in Fig. 13, $\Gamma_k = (i,u,v,s)$. Although different geometric objects can be considered for establishing the path of a musculotendon unit, the final result at each instant is composed of a set of points; and, the position vector for each of these points is known with respect to the local coordinate frames. As depicted in Fig. 13, a special notation has been used to differentiate between these points. For instance, $P_{ki}^{n}$ is the $n^{th}$ point of the $k^{th}$ musculotendon unit located on the $i^{th}$ body.

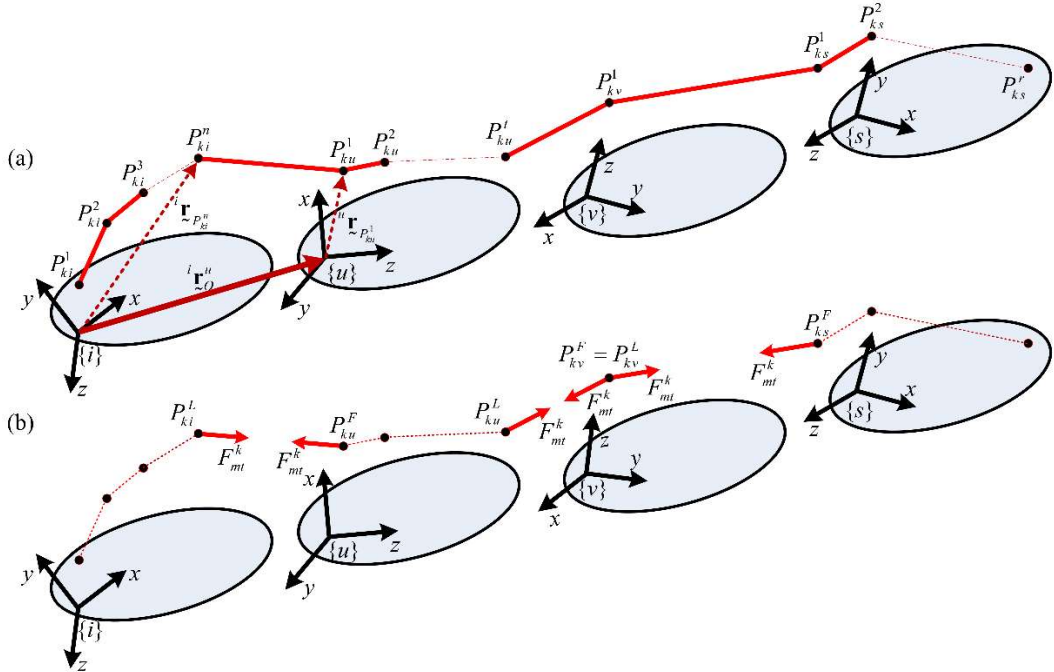

Figure 13 (a) A general musculotendon path passing through four bodies. As depicted in the figure, there is no inherent immediate association between the four bodies, such as being parent or ancestor. (b) The free body diagram illustrating the musculotendon unit corresponding to part (a). For a comprehensive understanding of the notation used, please refer to the accompanying text.

In order to specify the effect of a musculotendon unit on a body, its line of action vector should be determined. Considering the free body diagram illustrated in Fig. 13 (b), one may conclude that in order to obtain the line of action vector, knowing the position vector of all the points that assemble the path of the musculotendon unit is not necessary. For instance, in order to determine the line of action of $k^{th}$ musculotendon unit with respect to the origin body, i.e. the $i^{th}$ body in Fig. 13, one may need $P_{ki}^n$ and $P_{kv}^1$. These two points are the effective points of the corresponding line of action vector. Since in establishing the line of action vector, only the first and/or last points are involved, these two points are indicated as $P_{ki}^F$ and $P_{ki}^L$, respectively. In this notation, F and L refer to the "First" and "Last" and should not be mistaken as indices. Considering the $O_i P_{ki}^L P_{ku}^F O_u$ quadrilateral, one can write a loop equation as follows:

$$^i\mathbf{r}_{P_{ki}^L} + \mathbf{r}_{P_{ki}^L P_{ku}^F} = {}^i\mathbf{r}_O^u + {}^u\mathbf{r}_{P_{ku}^F}. \tag{150}$$

Rearranging the terms in this equation and transforming the $u^{th}$ coordinate bases into the $i^{th}$ one, yields

$$\mathbf{r}_{P_{ki}^L P_{ku}^F} = {}^i\mathbf{r}_O^u + {}^i_u\mathbf{A}\,{}^u\mathbf{r}_{P_{ku}^F} - {}^i\mathbf{r}_{P_{ki}^L}. \tag{151}$$

In other words,

$$^i\mathbf{e}_{mt}^k = \frac{{}^{(i)}\mathbf{r}_{P_{ki}^L P_{ku}^F}}{\left\|{}^{(i)}\mathbf{r}_{P_{ki}^L P_{ku}^F}\right\|} \tag{152}$$

where, $^i\mathbf{e}_{mt}^k$ is the line of action vector for the $k^{th}$ musculotendon unit with respect to the $i^{th}$ body and $\|.\|$ depicts 2-norm for vectors. A reciprocal line of action vector, i.e., $^i\bar{\mathbf{e}}_{mt}^k$, is defined as $^i\bar{\mathbf{e}}_{mt}^k = -{}^u_i\mathbf{A}\,{}^i\mathbf{e}_{mt}^k$. By considering every two consecutive effective points and employing the same procedure, one may obtain the corresponding line of action vector; however, using the concept of reciprocal line of action vector, redundant computations are avoided. Since the effect of the $k^{th}$ musculotendon unit on the body $u$ is determined at two points (see Fig. 13(b)), the idea of line of action may seem different from what was discussed for the $i^{th}$ body. In order to generalize the concept of the line of action vector in the sequel, I shall define the "Generalized Musculotendon Line of Action Vector" (GMLAV).

## Definition 3 – Generalized Musculotendon Line of Action Vector (GMLAV)

Assuming $j^* = \Gamma_k(j)$ and provided that $\Gamma_k(j-1)$ exists, then $j^{**} = \Gamma_k(j-1)$, I define

$$_{(j^*)}\mathbf{E}_{kj^*} = \begin{cases} {}^{j^*}\mathbf{e}_{mt}^k & j = 1 \\ {}^{j^*}\mathbf{e}_{mt}^k + {}^{j^{**}}\overline{\mathbf{e}}_{mt}^k & j = 2, 3, \ldots, N(\Gamma_k) - 1 \\ {}^{j^{**}}\overline{\mathbf{e}}_{mt}^k & j = N(\Gamma_k) \end{cases}$$

where, $_{(j^*)}\mathbf{E}_{kj^*}$ is the generalized musculotendon line of action for the $k^{th}$ muscle with respect to the $j^{*th}$ body. ${}^{j^{**}}\overline{\mathbf{e}}_{mt}^k$ (the reciprocal line of action vector) is defined as ${}^{j^{**}}\overline{\mathbf{e}}_{mt}^k = -{}_{j^{**}}^{j^*}\mathbf{A}\, {}^{j^*}\mathbf{e}_{mt}^k$; and, $N(\cdot)$ is an integer-valued function which provides the cardinality of a sequence.

In this dissertation, the GMLAV concept is utilized to define the force vector of musculotendon units. As the upcoming paragraphs involve the transformation of these vectors into the origin of the local coordinate frames, it becomes crucial to establish the moment of GMLAV, which is defined as follows:

## Definition 4 – Moment of GMLAV

Considering the same assumptions as Definition 3, I introduce,

$$_{(j^*)}\hat{\mathbf{E}}_{kj^*} = \begin{cases} {}^{j^*}\tilde{\mathbf{r}}_{P_{kj^*}^L}\, {}^{j^*}\mathbf{e}_{mt}^k & j = 1 \\ {}^{j^*}\tilde{\mathbf{r}}_{P_{kj^*}^L}\, {}^{j^*}\mathbf{e}_{mt}^k + {}^{j^{**}}\tilde{\mathbf{r}}_{P_{kj^*}^F}\, {}^{j^{**}}\overline{\mathbf{e}}_{mt}^k & j = 2, 3, \ldots, N(\Gamma_k) - 1 \\ {}^{j^{**}}\tilde{\mathbf{r}}_{P_{kj^*}^F}\, {}^{j^{**}}\overline{\mathbf{e}}_{mt}^k & j = N(\Gamma_k) \end{cases}$$

where $_{(j^*)}\hat{\mathbf{E}}_{kj^*}$ is the moment of GMLAV for the $k^{th}$ musculotendon unit with respect to the $j^{*th}$ body. Other quantities and notations have been defined earlier.

In the previously mentioned definitions, the GMLAV was only considered for the bodies that are part of the attachment sequence of a particular musculotendon unit. To further expand on this concept, Definition 5 is introduced, which enables the determination of the GMLAV (and moment of GMLAV) of a specific musculotendon unit with respect to all the bodies comprising the system.

**Definition 5 – Actuator Membership Set for ith body**

$\Lambda_i$ is a set that contains all the indices of the musculotendon units that exert force on the $i^{th}$ body and is defined as follows:

$$\Lambda_i = \left\{ k \mid k \in \{1, 2, \ldots, n_{mt}\} \ \& \ k^{th} \text{ musculotendon unit acts upon body } i \right\}$$

Considering this definition, the GMLAV of the $k^{th}$ musculotendon unit with respect to an arbitrary body $i$ is defined as

$$_{(i)}\mathbf{E}_{ki} = \begin{cases} _{(i)}\mathbf{E}_{k\Gamma_k(j)} & k \in \Lambda_i \\ \mathbf{0}_{3\times 1} & k \notin \Lambda_i \end{cases}, \tag{153}$$

provided that $k \in \Lambda_i$, $i = \Gamma_k(j)$.

Similarly, the moment of GMLAV is

$$_{(i)}\hat{\mathbf{E}}_{ki} = \begin{cases} _{(i)}\hat{\mathbf{E}}_{k\Gamma_k(j)} & k \in \Lambda_i \\ \mathbf{0}_{3\times 1} & k \notin \Lambda_i \end{cases}. \quad (154)$$

Based on Eq. 153 and Eq. 154, all musculotendon units acting upon the $i^{th}$ body are transmitted to the origin of this body. Consequently, the resultant force becomes

$$_{(i)}\mathbf{F}_O^i = \sum_{k=1}^{n_{mt}} F_{mt}^j \,_{(i)}\mathbf{E}_{ki}. \quad (155)$$

This equation can be reformulated in the following matrix format:

$$_{(i)}\mathbf{F}_o^i = \,_{(i)}\mathbf{E}_i^T \mathbf{F}_{mt} \quad (156)$$

where $\mathbf{E}_i \in \mathbb{R}^{n_{mt}\times 3}$ and is defined as

$$\mathbf{E}_i = \begin{Bmatrix} \mathbf{E}_{1i}^T \\ \mathbf{E}_{2i}^T \\ \vdots \\ \mathbf{E}_{n_{mt}i}^T \end{Bmatrix}. \quad (157)$$

The resultant moment at the origin, i.e., $\mathbf{M}_o^i$, becomes

$$_{(i)}\mathbf{M}_o^i = \sum_{k=1}^{n_{mt}} F_{mt}^j \,_{(i)}\hat{\mathbf{E}}_{ki}. \quad (158)$$

Also,

$$_{(i)}\mathbf{M}_o^i = \,_{(i)}\hat{\mathbf{E}}_i^T \mathbf{F}_{mt} \quad (159)$$

where

$$\hat{\mathbf{E}}_i = \begin{Bmatrix} \hat{\mathbf{E}}_{1i}^T \\ \hat{\mathbf{E}}_{2i}^T \\ \vdots \\ \hat{\mathbf{E}}_{n_{mt}i}^T \end{Bmatrix}. \tag{160}$$

Next, the virtual work associated with $_{(i)}\mathbf{F}_O^i$ and $_{(i)}\mathbf{M}_O^i$ is calculated as follows:

$$\delta W_{\text{MT}}^i = \left[_{(i)}\left(\delta^0\mathbf{r}_O^i\right)\right]^T {}_{(i)}\mathbf{F}_O^i + \left[_{(i)}\left(\delta^0\Theta^i\right)\right]^T {}_{(i)}\mathbf{M}_O^i \tag{161}$$

where, $\delta^0\mathbf{r}_O^i$ is the virtual position vector of the origin of the $i^{th}$ coordinate frame with respect to the inertial one; and, $\delta^0\Theta^i$ is the virtual orientation vector of the $i^{th}$ body with respect to the inertial reference frame. In what follows I shall elucidate on how to obtain these two quantities using the calculus of matrices; then, to derive the moment arm matrix of the musculotendon units I will expand the analysis to the entire multibody system.

A) Deriving $\left[_{(i)}\left(\delta^0\mathbf{r}_O^i\right)\right]^T$

Using $\begin{Bmatrix} {}^0\mathbf{r}_O^i \\ \hline 1 \end{Bmatrix} = {}_i^0\mathbf{A}_4\overline{\mathbf{0}}$, yields

$$\begin{Bmatrix} \delta^0\mathbf{r}_O^i \\ \hline 0 \end{Bmatrix} = \left(\delta_i^0\mathbf{A}_4\right)\overline{\mathbf{0}}; \tag{162}$$

consequently,

$$\begin{Bmatrix} {}_{(i)}\left(\delta^0\mathbf{r}_O^i\right) \\ \hline 0 \end{Bmatrix} = {}_0^i\overline{\mathbf{A}}_4\left(\delta_i^0\mathbf{A}_4\right)\overline{\mathbf{0}} \tag{163}$$

where,

$$_0^i\bar{\mathbf{A}}_4 = \left[\begin{array}{c|c} _0^i\mathbf{A} & \mathbf{0}_{3\times 1} \\ \hline \mathbf{0}_{3\times 1}^T & 1 \end{array}\right]. \tag{164}$$

Using part b of Theorem 4 results in

$$\delta\,_i^0\mathbf{A}_4 = \left(\frac{\partial \mathbf{B}_i}{\partial \mathbf{q}}\right)^T (\mathbf{I}_4 \otimes \delta\mathbf{q}). \tag{165}$$

As mentioned before, for notational conciseness I have assumed that $\mathbf{B}_i = {}_i^0\mathbf{A}_4^T$. Substituting this equation into Eq. 163 yields

$$\left\{\frac{\delta\left({}^0\mathbf{r}_O^i\right)}{0}\right\} = {}_0^i\bar{\mathbf{A}}_4 \left(\frac{\partial \mathbf{B}_i}{\partial \mathbf{q}}\right)^T (\mathbf{I}_4 \otimes \delta\mathbf{q})\bar{\mathbf{0}}. \tag{166}$$

Transposing both sides of this relation results in

$$\left\{\left[{}_{(i)}\left(\delta^0\mathbf{r}_O^i\right)\right]^T \;\middle|\; 0\right\} = \bar{\mathbf{0}}^T (\mathbf{I}_4 \otimes \delta\mathbf{q}^T)\left(\frac{\partial \mathbf{B}_i}{\partial \mathbf{q}}\right){}_i^0\bar{\mathbf{A}}_4. \tag{167}$$

Applying the trivial identity $\bar{\mathbf{0}}^T = \bar{\mathbf{0}}^T \otimes 1$, and using the mixed-product property of the Kronecker multiplication, this equation is rearranged as follows:

$$\left\{\left[{}_{(i)}\left(\delta^0\mathbf{r}_O^i\right)\right]^T \;\middle|\; 0\right\} = (\bar{\mathbf{0}}^T \otimes \delta\mathbf{q}^T)\left(\frac{\partial \mathbf{B}_i}{\partial \mathbf{q}}\right){}_i^0\bar{\mathbf{A}}_4. \tag{168}$$

Simplifying this equation yields

$$\left\{\left[{}_{(i)}\left(\delta^0\mathbf{r}_O^i\right)\right]^T \;\middle|\; 0\right\} = \delta\mathbf{q}^T \left(\bar{\mathbf{0}}^T \otimes \mathbf{I}_{n_q}\right)\left(\frac{\partial \mathbf{B}_i}{\partial \mathbf{q}}\right){}_i^0\bar{\mathbf{A}}_4; \tag{169}$$

in other words,

$$\left\{\left[{}_{(i)}\left(\delta^0\mathbf{r}_O^i\right)\right]^T \;\middle|\; 0\right\} = \delta\mathbf{q}^T \gamma_i^* \tag{170}$$

where $\gamma_i^* \in \mathbb{R}^{n_q \times 4}$, and is defined as

$$\gamma_i^* = \left(\overline{\mathbf{0}}^T \otimes \mathbf{I}_{n_q}\right)\left(\frac{\partial \mathbf{B}_i}{\partial \mathbf{q}}\right){}_i^0\overline{\mathbf{A}}_4 . \tag{171}$$

Substituting Eq. 75 into the right-hand side of this equation results in

$$\gamma_i^* = \left(\overline{\mathbf{0}}^T \otimes \mathbf{I}_{n_q}\right)\left(\sum_{j \in Anc(i) \cup \{i\}} \mathbf{W}_{ji}^T \otimes \nabla f_j\right){}_i^0\overline{\mathbf{A}}_4 . \tag{172}$$

Applying the mixed-product property of the Kronecker multiplication on this equation yields

$$\gamma_i^* = \sum_{j \in Anc(i) \cup \{i\}} \left(\overline{\mathbf{0}}^T \mathbf{W}_{ji}^T {}_i^0\overline{\mathbf{A}}_4\right) \otimes \nabla f_j . \tag{173}$$

Finally, using the definition of ${}_i^0\overline{\mathbf{A}}_4$ and $\mathbf{W}_{ji}$, one can draw the following conclusion

$$\overline{\mathbf{0}}^T \mathbf{W}_{ji}^T {}_i^0\overline{\mathbf{A}}_4 = \begin{cases} \left\{\left({}^j\mathbf{r}_O^i\right)^T \tilde{\mathbf{V}}_{\theta_j}^T {}_i^j\mathbf{A} \mid 0\right\} & j \equiv \text{Revolute} \\ \left\{\left({}_{pr(j)}^j\mathbf{A}\mathbf{V}_{d_j}\right)^T {}_i^j\mathbf{A} \mid 0\right\} & j \equiv \text{Prismatic} \end{cases} . \tag{174}$$

Using this important result, the following is immediately concluded

$$\left[{}_{(i)}\left(\delta {}^0\mathbf{r}_O^i\right)\right]^T = \delta \mathbf{q}^T \gamma_i \tag{175}$$

where

$$\gamma_i = \sum_{j \in Anc(i) \cup \{i\}} \begin{cases} \left[\left({}^j\mathbf{r}_O^i\right)^T \tilde{\mathbf{V}}_{\theta_j}^T {}_i^j\mathbf{A}\right] \otimes \nabla f_j & j \equiv \text{Revolute} \\ \left[\left({}_{pr(j)}^j\mathbf{A}\mathbf{V}_{d_j}\right)^T {}_i^j\mathbf{A}\right] \otimes \nabla f_j & j \equiv \text{Prismatic} \end{cases} . \tag{176}$$

B) Deriving $\left[{}_{(i)}\left(\delta {}^0\Theta^i\right)\right]^T$

The angular velocity vector of the $i^{th}$ body with respect to the inertial reference frame, i.e., $^0\boldsymbol{\omega}^i$, is

$$_{(i)}^{0}\boldsymbol{\omega}^i = \sum_{j=1}^{n_b} \vartheta_{j\,(i)}\boldsymbol{\beta}_{ji}^T \left(\nabla f_j\right)^T \dot{\mathbf{q}}, \tag{177}$$

where $\vartheta_j$ is a binary variable which determines the joint type

$$\vartheta_j = \begin{cases} 1 & j \equiv \text{Revolute} \\ 0 & j \equiv \text{Prismatic} \end{cases} \tag{178}$$

besides,

$$_{(i)}\boldsymbol{\beta}_{ji}^T = \begin{cases} _{pr(j)}^{i}\mathbf{AV}_{\theta_j} & j \in Anc(i) \cup \{i\} \\ \mathbf{0}_{3\times 1} & \text{otherwise} \end{cases}. \tag{179}$$

Taking the definition of the Jacobian matrix into consideration, the following can be derived:

$$_{(i)}^{0}\boldsymbol{\omega}^i = {}_{(i)}\boldsymbol{\beta}_i^T \vartheta \mathbf{J} \dot{\mathbf{q}} \tag{180}$$

where $\vartheta$ is a diagonal matrix defined as

$$\vartheta = diag(\vartheta_1, \vartheta_2, \ldots, \vartheta_{n_b}). \tag{181}$$

Also $\boldsymbol{\beta}_i \in \mathbb{R}^{n_b \times 3}$, and is defined as

$$\boldsymbol{\beta}_i = \begin{Bmatrix} \boldsymbol{\beta}_{1i} \\ \boldsymbol{\beta}_{2i} \\ \vdots \\ \boldsymbol{\beta}_{n_b i} \end{Bmatrix}. \tag{182}$$

Given the relationship between the rotation tuple and the angular velocity vector, one can extract $\delta^0\Theta^i$ from Eq. 180 as follows:

$$_{(i)}\left(\delta^0\Theta^i\right) = {}_{(i)}\boldsymbol{\beta}_i^T \boldsymbol{\vartheta} \mathbf{J} \, \delta\mathbf{q} \, . \tag{183}$$

Consequently,

$$\left[{}_{(i)}\left(\delta^0\Theta^i\right)\right]^T = \delta\mathbf{q}^T \mathbf{J}^T \boldsymbol{\vartheta} \, {}_{(i)}\boldsymbol{\beta}_i \, . \tag{184}$$

Substituting Eq. 175 and Eq. 184 into Eq. 161, the virtual work of the musculotendon units acting upon the $i^{th}$ body is derived

$$\delta W_{MT}^i = \delta\mathbf{q}^T \left({}_{(i)}\boldsymbol{\gamma}_i \, {}_{(i)}\mathbf{E}_i^T + \mathbf{J}^T \boldsymbol{\vartheta} \, {}_{(i)}\boldsymbol{\beta}_i \, {}_{(i)}\hat{\mathbf{E}}_i^T\right)\mathbf{F}_{mt} \, . \tag{185}$$

### 2-2-4 Virtual work of the weight force

The weight force of the $i^{th}$ body is described as $m_i \, {}_{(0)}\mathbf{g}$; where ${}_{(0)}\mathbf{g}$ is the vector of acceleration due to gravity, and $m_i$ is the mass of the $i^{th}$ body. Using Eq. 46, the virtual work corresponding to the weight force of the $i^{th}$ body is

$$\delta W_{mg}^i = \delta\mathbf{q}^T \mathbf{G}_i \tag{186}$$

where,

$$\mathbf{G}_i = m_i \left(\left\{{}^i\mathbf{r}_{cm}^T \mid 1\right\} \otimes \mathbf{I}_{n_q}\right) \mathbf{D}_i \left\{\begin{matrix} {}_{(0)}\mathbf{g} \\ 0 \end{matrix}\right\} . \tag{187}$$

### 3-4 The virtual work principle for the multibody

Based on the preceding results, the inertial virtual work corresponding to the entire multibody system is determined as

$$\delta W_{\mathrm{I}} = -\delta \mathbf{q}^T \left( \mathbf{M} \ddot{\mathbf{q}} + \mathbf{N} \dot{\mathbf{q}} \right) \tag{188}$$

where,

$$\mathbf{M} = \sum_{i=1}^{n_b} \mathbf{M}_i \tag{189}$$

and,

$$\mathbf{N}_i = \sum_{i=1}^{n_b} \mathbf{N}_i . \tag{190}$$

Also, the musculotendon virtual work corresponding to the entire multibody is derived as

$$\delta W_{\mathrm{MT}} = \delta \mathbf{q}^T \left( \sum_{i=1}^{n_b} {}_{(i)}\boldsymbol{\gamma}_{i\ (i)} \mathbf{E}_i^T + \mathbf{J}^T \boldsymbol{\vartheta}_{\ (i)} \boldsymbol{\beta}_{i\ (i)} \hat{\mathbf{E}}_i^T \right) \mathbf{F}_{mt} . \tag{191}$$

The equation is reformulated as

$$\delta W_{\mathrm{MT}} = \delta \mathbf{q}^T \mathbf{R} \mathbf{F}_{mt}, \tag{192}$$

where $\mathbf{R}$ is the moment arm matrix associated with the musculotendon units, and is defined as

$$\mathbf{R} = \boldsymbol{\gamma}^T \mathbf{E} + \mathbf{J}^T \boldsymbol{\vartheta} \boldsymbol{\beta}^T \hat{\mathbf{E}}, \tag{193}$$

where, $\boldsymbol{\gamma}$, $\mathbf{E}$, $\boldsymbol{\beta}$ and $\hat{\mathbf{E}}$ are block matrices defined in the following:

$$\boldsymbol{\gamma} = \begin{bmatrix} {}_{(1)}\boldsymbol{\gamma}_1^T \\ \hline {}_{(2)}\boldsymbol{\gamma}_2^T \\ \hline \vdots \\ \hline {}_{(n_b)}\boldsymbol{\gamma}_{n_b}^T \end{bmatrix}, \mathbf{E} = \begin{bmatrix} {}_{(1)}\mathbf{E}_1^T \\ \hline {}_{(2)}\mathbf{E}_2^T \\ \hline \vdots \\ \hline {}_{(n_b)}\mathbf{E}_{n_b}^T \end{bmatrix}, \boldsymbol{\beta} = \begin{bmatrix} {}_{(1)}\boldsymbol{\beta}_1^T \\ \hline {}_{(2)}\boldsymbol{\beta}_2^T \\ \hline \vdots \\ \hline {}_{(n_b)}\boldsymbol{\beta}_{n_b}^T \end{bmatrix}, \hat{\mathbf{E}} = \begin{bmatrix} {}_{(1)}\hat{\mathbf{E}}_1^T \\ \hline {}_{(2)}\hat{\mathbf{E}}_2^T \\ \hline \vdots \\ \hline {}_{(n_b)}\hat{\mathbf{E}}_{n_b}^T \end{bmatrix} . \tag{194}$$

Although Eq. 193 provides a closed-form formula for the moment arm matrix of the musculotendon units, this equation can be reformulated to enhance computational efficiency. As stated before, the GMLAV of the $k^{th}$ musculotendon unit with respect to an arbitrary body is the zero vector except for the bodies that belong to the attachment sequence of this musculotendon unit. Expanding and scrutinizing the block matrix $\mathbf{E}$, one may realize that the $k^{th}$ column of this matrix, i.e. $\mathbf{E}_{.,k}$, consists of all the GMLAV for the $k^{th}$ musculotendon unit. This column is composed of $3 \times 1$ blocks; each block depicts the GMLAV of the $k^{th}$ musculotendon unit with respect to a specific body. Therefore, due to the nature of a general musculoskeletal system, i.e., one muscle typically only spans several bodies, an extensive number of zeros inherently exists in this column. Considering the definition of the moment of GMLAV, it is straightforward to infer the same topology for $\hat{\mathbf{E}}$. Leveraging this information, I shall enhance the computational efficiency of Eq. 193. Since the only non-zero blocks of $\mathbf{E}_{.,k}$ and $\hat{\mathbf{E}}_{.,k}$ belong to the attachment sequence of the $k^{th}$ musculotendon unit, instead of computing $\boldsymbol{\gamma}^T \mathbf{E}_{.,k}$ and $\boldsymbol{\beta}^T \hat{\mathbf{E}}_{.,k}$ it suffices to consider the linear combination of the non-zero blocks with their corresponding gamma and beta coefficients. Therefore:

$$\mathbf{R}_{.,k} = \sum_{i \in \Gamma_k - \{0\}} \boldsymbol{\gamma}_i \mathbf{E}_{ki} + \mathbf{J}^T \boldsymbol{\vartheta} \sum_{i \in \Gamma_k - \{0\}} \boldsymbol{\beta}_i \hat{\mathbf{E}}_{ki} \tag{195}$$

where $\mathbf{R}_{.,k}$ denotes the $k^{th}$ column of the moment arm matrix.

Considering the entire multibody, the virtual work corresponding to the weight force is derived as

$$\delta W_{mg} = \delta \mathbf{q}^T \mathbf{G} \tag{196}$$

where

$$\mathbf{G} = \sum_{i=1}^{n_b} \mathbf{G}_i . \tag{197}$$

Finally, using the virtual work principle, the differential equations governing the entire musculoskeletal system are established

$$\mathbf{M}(\mathbf{q})\ddot{\mathbf{q}} + \mathbf{N}(\mathbf{q},\dot{\mathbf{q}})\dot{\mathbf{q}} = \mathbf{G}(\mathbf{q}) + \mathbf{R}(\mathbf{q})\mathbf{F}_{mt} \tag{198}$$

In what follows, three mechanisms will be simulated using this formalism.

## 4-4 Simulations

### A) Moment arm matrix formula of the musculotendon units[6]

In this subsection, the focus will be on the previously developed model by Holzbaur et al. (Holzbaur, Murray et al. 2005) for the shoulder mechanism, along with proposed method to calculate the moment arm matrix of the musculotendon units. Holzbaur et al. utilized empirical data from (de Groot and Brand 2001) to model the shoulder rhythm. However, certain simplifications were made in the empirical data when implementing the model in SIMM.

In the original empirical data, the orientation of the clavicle and scapula were represented as linear functions of the elevation angle, as well as the plane of elevation. However, since SIMM used to consider each joint as a function of a single generalized coordinate, Holzbaur et al. disregarded the influence of the plane of elevation on the orientation of the clavicle and scapula.

---

[6] This simulation was previously published in Ehsani, H., M. Rostami and M. Parnianpour (2016). "A closed-form formula for the moment arm matrix of a general musculoskeletal model with considering joint constraint and motion rhythm." Multibody System Dynamics 36: 377-403.

The formalism presented in this dissertation effectively addresses the aforementioned limitation. However, to enable a comprehensive comparison and evaluation, a similar shoulder model employed by Holzbaur et al. will be utilized in this section. The moment arm results obtained from our proposed formulation will be compared to those obtained from OpenSim.

To ensure an appropriate comparison with the outcomes of OpenSim, the musculotendon units considered in Holzbaur's model are categorized into groups. From each group, one unit is designated as the leader and included in the collation process. This categorization, as presented in Table 5, is based on the attachment sequence of the musculotendon units. Additionally, to accurately define the path of certain musculotendon units like the anterior deltoid, Holzbaur introduced moving via points. These points were incorporated into SIMM as bodies with zero inertia, articulating them to the respective segments (Holzbaur 2005). This consideration of moving via points is taken into account during the categorization of the musculotendon units spanning the shoulder.

To calculate the moment arms, all generalized coordinates of the model are systematically varied within their range of motion using a 1-degree step length. For example, the elevation angle is adjusted from 0 to 180 degrees with a 1-degree increment, and the moment arms are computed for each generalized coordinate. In this analysis, the remaining two generalized coordinates, namely the elevation plane and axial rotation, are assumed to be in the anatomical zero position.

Since three generalized coordinates are incorporated into this model, to compare the results with those obtained from OpenSim, 9 different conditions can be considered. To showcase the results, I shall present two of these conditions in what follows. In Fig. 14, the moment arms of the leader of each group (see Table 5) with respect to the second generalized coordinate of the model (i.e., elevation angle) is plotted against the elevation angle ($q_1$). The results obtained from the closed-form formula presented in this dissertation (solid lines) are juxtaposed with those obtained from OpenSim (dashed lines). Similarly, in Fig. 15, the moment arm of the leaders with respect to the third generalized coordinate are plotted against the second generalized coordinate. In both cases, the results are identical to 8 significant figures.

Table 5 Categorization for the musculotendon units of the shoulder mechanism based on the model outlined in Holzbaur (Holzbaur 2005). The details of this categorization are explained in the body of the text.

| Group Id | Description (Attachment sequence) | Leader abbreviation | Members |
|---|---|---|---|
| 1 | Thorax-Scapula-MScapula1- MScapula2-Humerus | lat1 | Latissimus dorsi (Thoracic, lumbar and iliac compartments) |
| 2 | Thorax-MThorax-MHumerus-Humerus | pecm3 | Pectoralis major (Ribs and sternal compartments) |
| 3 | Clavicle-Scapula-Humerus | delt1 | Deltoid (Anterior compartment) |
| 4 | Clavicle-MThorax-MHumerus-Humerus | pecm1 | Pectoralis major (Clavicular compartment) |
| 5 | Scapula-Humerus | subsc | Subscapularis, Coracobrachilais, Teres major, Teres minor, Infraspinatus, Supraspinatus, Deltoid (Posterior compartment) |
| 6 | Scapula-MHumerus-Humerus | delt2 | Deltoid (Middle compartment) |

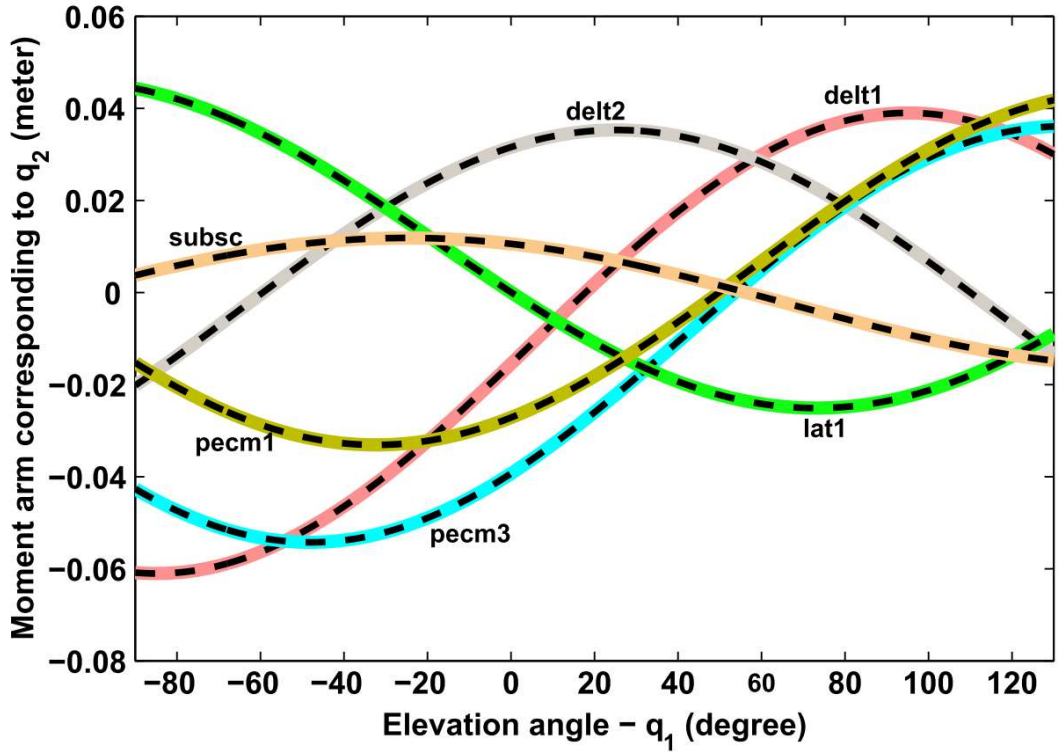

Figure 14 The moment arms of the group leaders of the shoulder mechanism corresponding to the shoulder elevation. These moment arms were obtained while the first generalized coordinate was altered through its entire range of motion from -90 to -130 degrees with 1 degree of resolution

## B) Andrew's Squeezer Mechanism[7]

In this section, the validity of the proposed formalism in deriving the differential equations of motion is examined through simulation of Andrew's squeezer mechanism. This particular mechanism exhibits a very small time constant, and when modeled using differential-algebraic equations, it leads to the phenomenon of drift-off. Consequently, obtaining a solution using conventional methods becomes challenging. As a result,

---

[7] Simulations of subsections B and C were previously published in Ehsani, H., M. Poursina, M. Rostami, A. Mousavi, M. Parnianpour and K. Khalaf (2019). "Efficient embedding of empirically-derived constraints in the ODE formulation of multibody systems: Application to the human body musculoskeletal system." Mechanism and Machine Theory 133: 673-690.

Andrew's squeezer mechanism is commonly regarded as a benchmark problem for evaluating new approaches in deriving equations of motion for multibody systems (Schiehlen 1990).

The planar mechanism depicted in Figure 16 consists of seven moving bodies connected to each other and to the ground through ideal revolute joints. It possesses one degree of freedom. The parameters of this model were adopted from (Hairer and Wanner 2010).

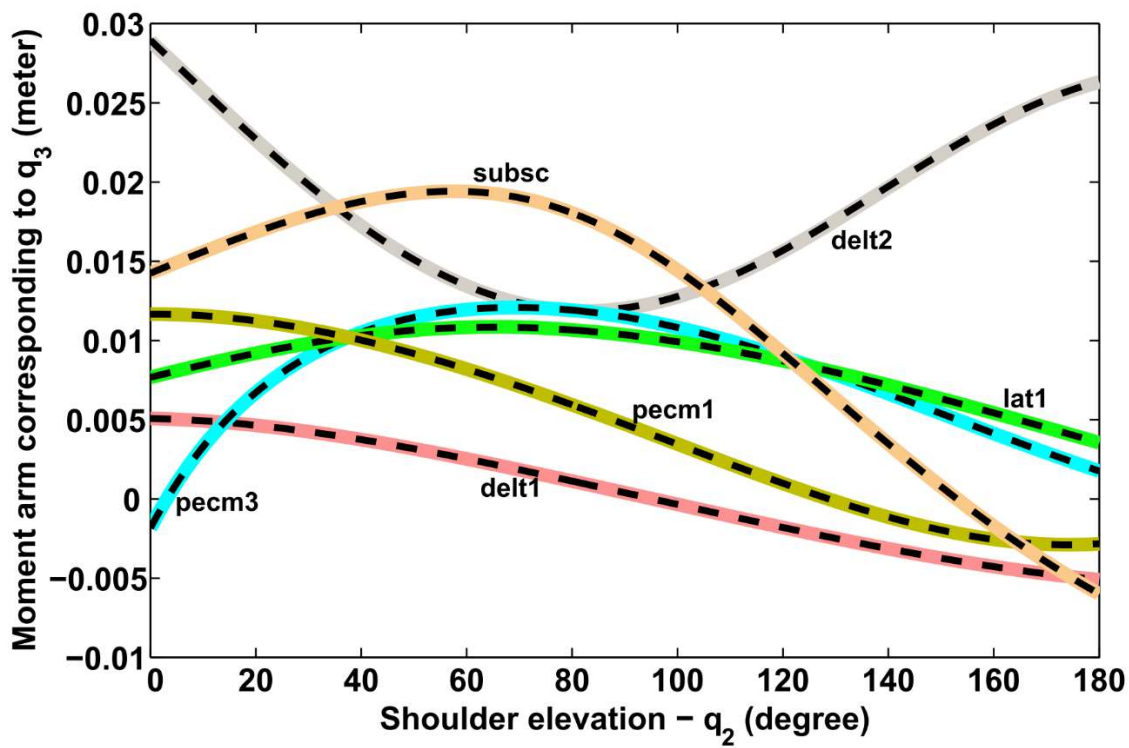

Figure 15 The moment arms of the group leaders of the shoulder mechanism corresponding to the shoulder rotation. These moment arms were obtained while the second generalized coordinate was altered through its entire range of motion from 0 to 180 degrees with 1 degree of resolution.

Considering the existing kinematic loops between the moving bodies and the fixed base, the following algebraic equations can be obtained between the joint variables ($\theta_i$ for

$i = 1, 2, ..., 7$) and the generalized coordinate ($q$):

$$\begin{cases} OA_x + AH\cos(\theta_2) + HE\cos(\theta_2 + \theta_6) + EF\cos(\theta_2 + \theta_6 + \theta_7) - OF\cos(\theta_1) = 0 \\ OA_y + AH\sin(\theta_2) + HE\sin(\theta_2 + \theta_6) + EF\sin(\theta_2 + \theta_6 + \theta_7) - OF\sin(\theta_1) = 0 \\ AG\cos(\theta_3) + GE\cos(\theta_3 + \theta_5) - AH\cos(\theta_2) - HE\cos(\theta_2 + \theta_6) = 0 \\ AG\sin(\theta_3) + GE\sin(\theta_3 + \theta_5) - AH\sin(\theta_2) - HE\sin(\theta_2 + \theta_6) = 0 \\ OA_x + AH\cos(\theta_2) + HE\cos(\theta_2 + \theta_6) - BE\cos(\theta_4) - OB_x = 0 \\ OA_y + AH\sin(\theta_2) + HE\sin(\theta_2 + \theta_6) - BE\sin(\theta_4) - OB_y = 0 \\ q - \theta_1 = 0 \end{cases} \quad (199)$$

where $OA_x$ and $OA_y$ denote the x and y coordinates of point A with respect to the inertial reference frame, respectively. A similar notation is adopted for the position components of point B. Other parameters and variables of Eq. 199 have been defined in Fig. 16.

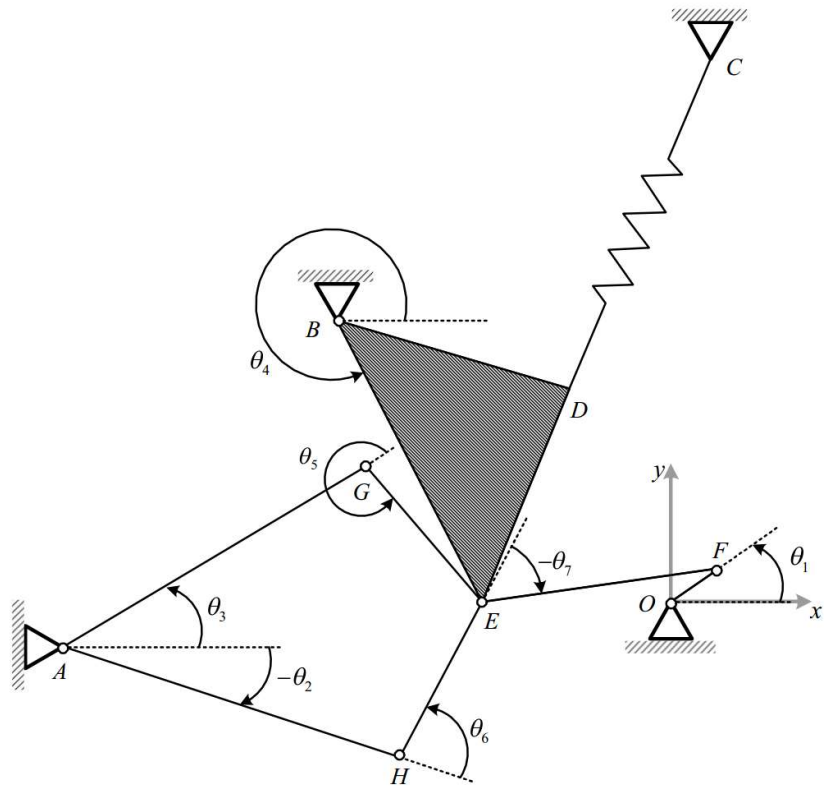

Figure 16 A schematic diagram of the considered Andrews' squeezer mechanism. The inertial reference frame is located at point O on the fixed base.

To begin, it is necessary to express the joint variables as smooth functions of the generalized coordinate. By incrementally changing the generalized coordinate over its entire range of motion with a step length of 0.01, the nonlinear algebraic equations (Eq. 199) are solved to determine the joint variables. These results provide the essential empirical information. Subsequently, utilizing Fourier approximations, the joint variables are mathematically expressed as functions of the generalized coordinate (refer to Fig. 17). Differentiating these smooth functions allows for obtaining the gradient vector and Hessian matrix for each joint variable. With this information and employing the formulation derived in the previous section, the differential equation representing the motion of this mechanism can be symbolically obtained in MATLAB.

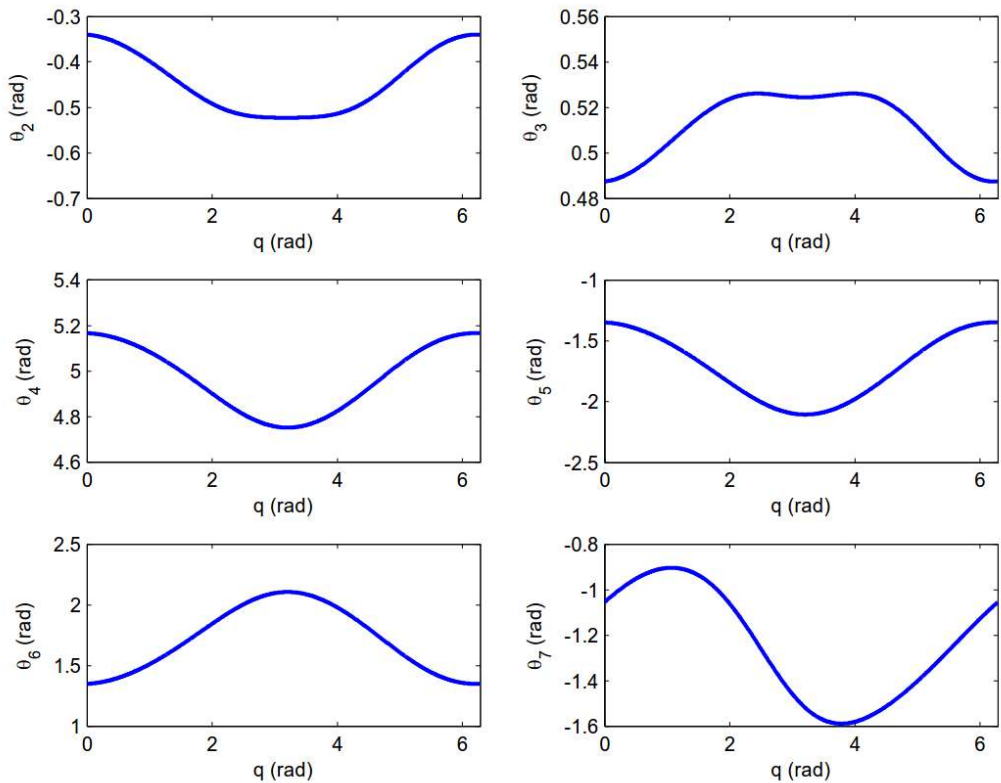

Figure 17 Joint variables of the Andrew's Squeezer Mechanism with respect to the generalized coordinate

Next, assuming appropriate initial conditions, such as $q=-0.0620\,\text{rad}$ and $\dot{q}=0\,\text{rad/s}$, the differential equation of motion is solved using the TR-BDF2 method (Bank, Coughran et al. 1985) for a time span of 0.05 seconds. Results are shown in Fig. 18.

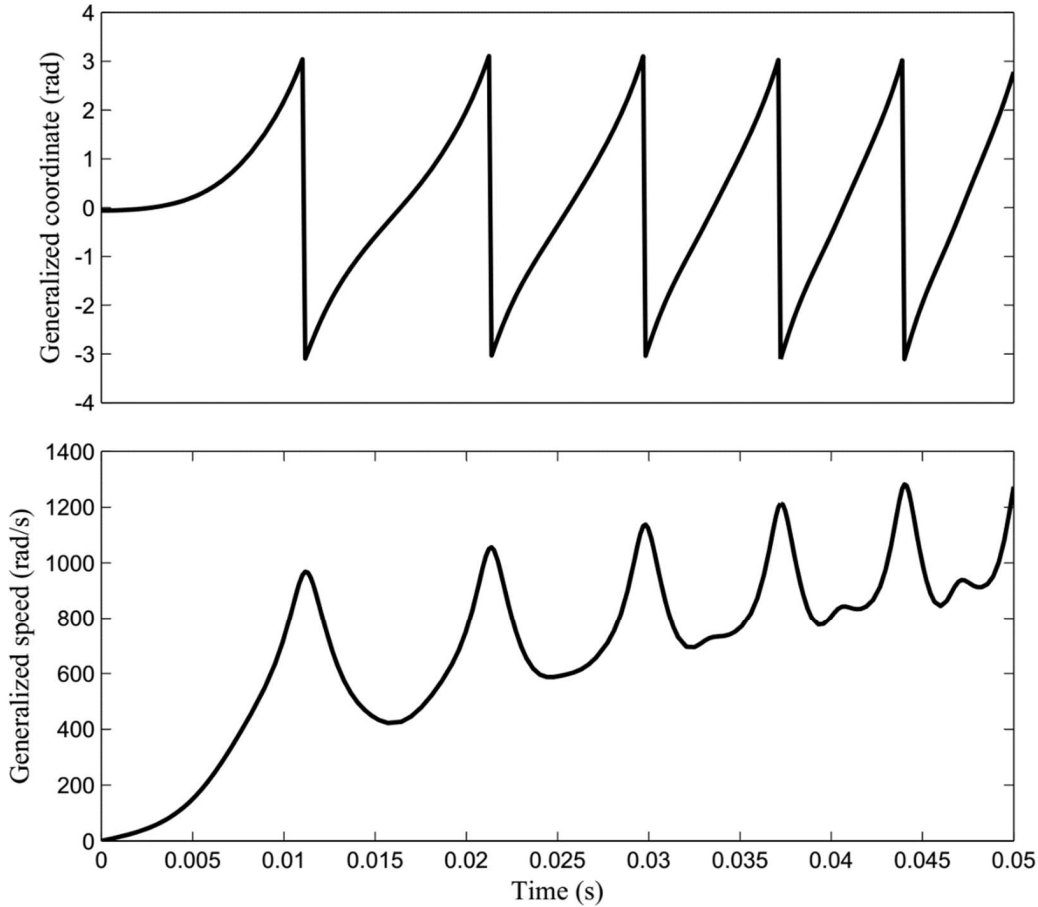

Figure 18 Generalized coordinate and generalized speed of the Anrew's Squeezer mechanism obtained from solving a forward dynamics problem using the method proposed in this dissertation

To further validate the results, a DAE model, as presented in (Hairer and Wanner 2010), is employed to simulate this mechanism. Consequently, the stated forward dynamics problem is also solved using an index reduction algorithm to obtain an index-I problem. To mitigate the drift-off phenomenon, this method is combined with the velocity stabilization technique. Using 1000 points, the RMS error between the results of these two methods is calculated. This error amounts to 0.0001 rad for the generalized coordinate and 0.003 rad/s for the generalized speed. I shall also compare the performance of these two methods in terms of simulation cost (time needed to integrate the equations). The empirically-based method was 4 times faster than the constraint stabilization technique in solving the stated forward dynamics problem. This computational efficiency mainly depends on the required step size to perform a stable numerical integration. In order to reach a stable solution, more nodes (i.e., smaller step size) are required to solve the DAE model, and therefore the simulation cost increases. In the empirically-based method, the algebraic constraints are embedded into the differential equations; therefore, the drift-off phenomenon does not occur, and hence no stabilization is required. As a result, a substantially smaller number of integration nodes is required for a stable solution.

**C) 3D Simulation of the Shoulder Mechanism**

In this subsection, using the model presented by de Groot and Brand for the shoulder rhythm, a motion tracking problem for the shoulder mechanism is solved (de Groot and Brand 2001). This model comprises the thorax as the fixed base, clavicle, and scapula. this model, the orientation of the clavicle and the scapula are contingent upon the plane of elevation and the elevation angle. The required Joints' axes, i.e. $\mathbf{V}_\theta$ and $\mathbf{V}_d$, are obtained from (Holzbaur 2005), while the anthropometric parameters, i.e. segment mass, location of the center of mass and moment of inertia tensor, are adopted from (Quental, Folgado et al. 2012).

The model is actuated with 14 musculotendon units: Deltoid (anterior, middle and posterior compartments), Subscapularis, Teres major, Teres minor, Infraspinatus, Supraspinatus, Pectoralis major (ribs, sternal and clavicular compartments) and Latissimusdorsi (thoracic, lumbar and iliac compartments). For modeling these actuators, a Hill-based phenomenological model is employed (Ehsani, Rostami et al. 2016).

As depicted in Fig. 19, a desired trajectory is considered for the elevation angle, while a fixed 90 degrees value and a -30 degrees value are assumed for the plane of elevation and shoulder axial rotation, respectively.

In order to determine the required torques ($\tau$) to generate this desired motion ($q_{des}$), a computed torque control method with PD controllers is employed as follows:

$$\begin{bmatrix} \mathbf{M} & -\mathbf{I}_3 \\ \mathbf{I}_3 & \mathbf{0}_{3\times 3} \end{bmatrix} \begin{Bmatrix} \ddot{\mathbf{q}} \\ \boldsymbol{\tau} \end{Bmatrix} = \begin{Bmatrix} \mathbf{G} - \mathbf{N} \\ \mathbf{Q}_d \end{Bmatrix} \tag{200}$$

and

$$\mathbf{Q}_d = \ddot{\mathbf{q}}_{des} - 2\mathbf{K}(\dot{\mathbf{q}} - \dot{\mathbf{q}}_{des}) - \mathbf{K}^2(\mathbf{q} - \mathbf{q}_{des}) \tag{201}$$

where $\mathbf{K}$ is a diagonal matrix containing the derivative controllers' gains. As shown in Eq. 201, the gains for the proportional controllers are designed to achieve a critically damped situation. Here, I shall consider K = diagonal.(100,100,1000).

Using the Runge-Kutta-Fehlberg method (Fehlberg 1969), Eq. 200 is solved for the state vectors, and subsequently the required torques are calculated (Fig. 20). For the first and the third generalized coordinates, the obtained results are identical to the corresponding desired trajectories up to 16 significant figures. The same is also true in the comparison of the generalized speeds. For the second generalized coordinates, the maximum discrepancy between the outcome of the tracking problem and the desired one is less than $2 \times 10^{-4}$ rad. Moreover, the maximum tracking error in the elevation angle was 0.02 rad/s.

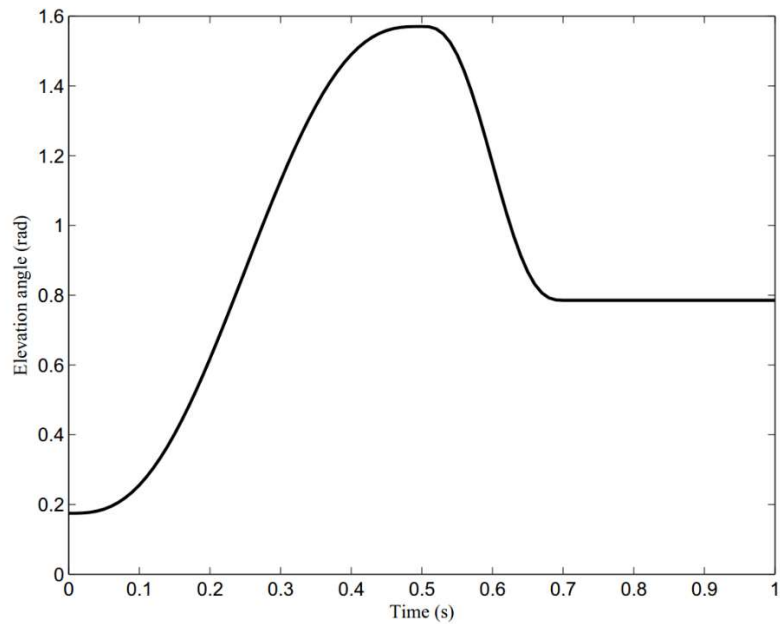

Figure 19 The desired trajectory for the elevation angle in the tracking problem

The human shoulder mechanism is also modeled via DAEs. Similar to the Andrews' squeezer mechanism, an index reduction algorithm is used to obtain Index I equations, and then solve the stated tracking problem with those equations. The accuracy of the results obtained from this method is similar to the outcome of the proposed method in this study. However, the simulation cost of the empirically-based method is significantly smaller, where the simulations performed more than 3 times faster as compared to DAE solution.

At the next stage, I shall use the torques to determine the forces exerted by the model's musculotendon units. Since the force-sharing problem is indeterminate (the number of unknowns is greater than the number of available equations), an optimization approach is considered. The squared sum of the activation levels of the muscles is considered as the objective function, while the balance equations between the calculated torques and the generalized tendon forces, and the governing equations of the musculotendon units are considered as constraints. The results in terms of the forces of the musculotendon units are shown in Fig 21.

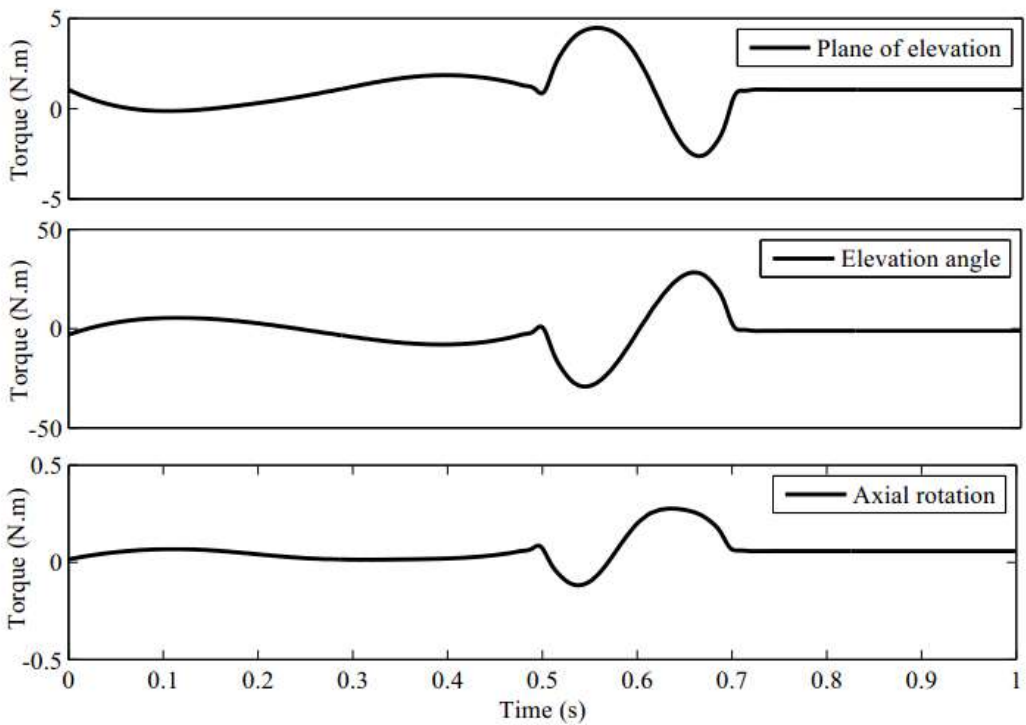

Figure 20 The required torques corresponding to the generalized coordinates of the shoulder model.

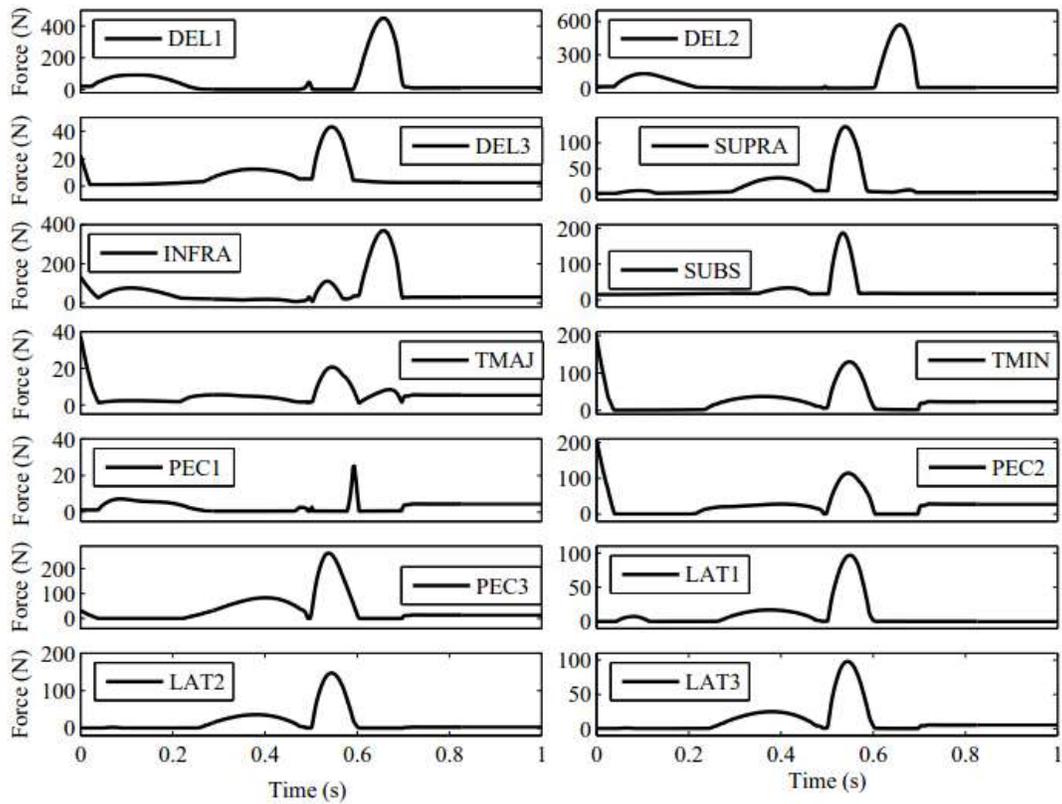

Figure 21 Time history diagrams of the forces of the musculotendon units of the shoulder model. The following abbreviations are adopted for the name of the muscles: Anterior deltoid (DEL1), Middle deltoid (DEL2), Posterior deltoid (DEL3), Supraspinatus (SUPRA), Infraspinatus (INFRA), Subscapularis (SUBS), Teres major (TMAJ), Teres minor (TMIN), Pectoralis major (clavicular (PEC1), ribs (PEC2) and sternal (PEC3)) and Latissimus dorsi (thoracic (LAT1, lumbar (LAT2) and iliac (LAT3)).